\documentclass[preprint,aps]{revtex4}
\usepackage{rotating}
\usepackage{verbatim}
\usepackage[normalem]{ulem}
\usepackage{bm} 
\usepackage{amssymb,amsmath,
amsthm,epsfig,subfigure} \newcommand{\ra}{\rangle} 
\newcommand{\la}{\langle}  
 \makeatletter  \newcommand{\Rmnum}[1]{\expandafter\@slowromancap\romannumeral #1@}

\newcommand{\be}{\begin{equation}}
\newcommand{\ee}{\end{equation}}
\newcommand{\ber}{\begin{eqnarray}}
\newcommand{\eer}{\end{eqnarray}}
\usepackage[belowskip=-30pt]{caption}
\makeatother

\usepackage{ulem}
\usepackage{color}

\begin{document}

\title{Time-dependent probability density functions and information geometry
in stochastic logistic and Gompertz models}
\author{Lucille-Marie Tenk\`es$^{1,3}$, Rainer Hollerbach$^2$ and Eun-jin Kim$^3$}
\affiliation{$^1$ENSTA ParisTech Universit\'e Paris-Saclay, 828, Boulevard des
Mar\'echaux 91120 Palaiseau, France\\
$^2$Department of Applied Mathematics, University of Leeds,
Leeds LS2 9JT, UK\\
$^3$School of Mathematics and Statistics, University of Sheffield,
Sheffield, S3 7RH, UK}

\vspace{1cm}

\begin{abstract}
A probabilistic description is essential for understanding growth processes far
from equilibrium. In this paper, we compute
time-dependent Probability Density Functions (PDFs) in order to investigate
stochastic logistic and Gompertz models, which are two of the most
popular growth models. We consider different types of short-correlated internal
(multiplicative) and external (additive) stochastic noises and compare the
time-dependent PDFs in the two models, elucidating the effects of the additive
and multiplicative noises on the form of PDFs. We demonstrate an interesting
transition from a unimodal to a bimodal PDF as the multiplicative noise
increases for a fixed value of the additive noise. A much weaker (leaky)
attractor in the Gompertz model leads to a significant (singular) growth of the
population of a very small size. We point out the limitation of using
stationary PDFs, mean value and variance in understanding statistical
properties of the growth far from equilibrium, highlighting the importance of
time-dependent PDFs. We further compare these two models from the
perspective of information change that occurs during the growth process.
Specifically, we define an infinitesimal distance at any time by comparing two
PDFs at times infinitesimally apart and sum these distances in time. The total
distance along the trajectory quantifies the total number of different states
that the system undergoes in time, and is called the information length. We
show that the time-evolution of the two models become more similar when
measured in units of the information length and point out the merit of 
using the information length in unifying and understanding the dynamic
evolution of different growth processes.
\end{abstract}

%\pacs{05.10.Gg,05.70.-a,89.70.Cf,47.10.Fg}
%{05.10.Gg -stochastic analysis
%05.70.Ln: Nonlinear equation, irreversible
%05.70.-a: thermodynamics
%89.70.Cf entropy, information
% 47.10.Fg: dynamical model for fluid turbulence
%52.35.a: plasma turbulence
\maketitle

\section{Introduction}

A variety of growth models are widely used in understanding the dynamic
evolution of populations, not only in humans, ecology and biosystems, but also
financial markets, environment, chemical and physical systems, etc. Two of the
most popular models are the logistic and Gompertz models
\cite{Tjorve2017,kimura64,AI03,ZHANG10,WANG11}. Mathematically, the utility of
these models stems from the simplicity in which the growth is saturated to a
finite value by a simple nonlinear damping term. Specifically, the logistic
model for the variable $x$ is governed by the following equation:
\begin{eqnarray}
\frac{dx}{dt} &=& \gamma x ( 1- \epsilon x),
\label{eq001}
\end{eqnarray}
which has the exact solution
\begin{eqnarray}
x(t) &=& \frac{x_0}{\epsilon x_0 +(1-\epsilon x_0) \exp{(-\gamma t)}}\,.
\label{eq002}
\end{eqnarray}
Here $\gamma$ and $\gamma \epsilon$ are positive constants, representing the
linear growth rate and nonlinear damping coefficient; $x_0$ is the value of $x$
at $t=0$. As $t$ becomes large, Eq.\ (\ref{eq002}) gives a stable solution
$x={1}/{\epsilon}$ which corresponds to the carrying capacity; $x=0$ is the
unstable equilibrium point. The linear term $\gamma x$ can be considered to
capture the overall effects (i.e.\ the birth rate minus death rate) that drive
the growth while the nonlinear damping term $-\gamma \epsilon x^{2}$ slows down
the growth and saturates $x$ to the finite equilibrium value given by the
carrying capacity $1/\epsilon$.\\

To compare the growth dynamics of $x$ in the logistic model (\ref{eq001}) with
that governed by the Gompertz model, in this paper we opt to use a different
variable $y$ as follows:
\begin{eqnarray}
\frac{dy}{dt} &=& \gamma y( 1 - \epsilon \ln{y}).
\label{eq003}
\end{eqnarray}
In comparison with the logistic model, the Gompertz model for the variable $y$
in Eq.\ (\ref{eq003}) has a similar linear growth term $\gamma y$ but a different
nonlinear damping $-\gamma \epsilon y \ln{y}$, which is weaker than the
quadratic damping in the logistic model. Eq.\ (\ref{eq003}) has the following
exact solution:
\begin{eqnarray}
y(t) &=& \exp \left[ \frac{1}{\epsilon} - \left( \frac{1}{\epsilon}
- \ln{y_0} \right) e^{-\epsilon \gamma t}\right],\label{eq004}
\end{eqnarray}
where $y_0$ is the value of $y$ at $t=0$. As $t$ becomes large, Eq.\ (\ref{eq004})
gives a stable solution $y=e^{{1}/{\epsilon}}$, and $y=0$ is the unstable
equilibrium point. The specific equilibrium points $x(t\to \infty)$ and
$y(t \to \infty)$ can easily be adjusted to take the same value in both models,
for example by taking $\epsilon=1/e$ for logistic and $\epsilon=1$ for Gompertz,
yielding $e$ as the stable solution in both models. However, the strength and
robustness of these equilibrium points in the presence of a stochastic noise can
differ significantly due to the different nonlinear damping terms, as we will
show in this paper.\\

The purpose of this paper is to investigate how these two models are affected
by stochastic noise, which is now believed to be crucial in many systems
\cite{SHALEK14,SCHENZLE79,MCADAMS97,SCHULTZ07,LEVY12,SIGAL06,YOKOTA00,DINGLI07,
FRANCK12,Buehrlng76,KIM08,KIM09,KIM13,tsuchiya15,Jensen98,Pruesser12,Longo11,
FLYNN2014,Ovidiu}.
For instance, variability has emerged to be a key factor in understanding the
development of tumours \cite{YOKOTA00,DINGLI07,FRANCK12}. Even when tumours may
have smaller growth rates compared to normal cells \cite{Buehrlng76}, its growth
will have significantly larger variation than normal cells. This variation
increases with the complexity of the tumour as it progresses towards metastasis
with the involvement of an increasing number of different processes on a broad
range of scales, accompanied by the propagation of the loss of cellular
stability on multiscales (e.g.\ \cite{YOKOTA00,DINGLI07,FRANCK12}). 
Ultimately, the danger of a tumour is not measured simply by its average growth
rate or variance, but by the rare occurrence of extreme events of such
aggressive tumour growth (metastasis). These extreme events are the
manifestation of intermittency, which is a generic feature in non-equilibrium
systems (e.g.\ \cite{KIM08,KIM13}).\\

In the presence of such stochasticity and intermittency, a probabilistic
description using a Probability Density Function (PDF), rather than relying
on mean value and variance, becomes essential. Such PDFs would also enable 
us to understand the fundamental mechanisms responsible for bimodal distributions 
(e.g.\ \cite{SHALEK14}). Experimentally, obtaining a good quality of PDFs is
often very challenging as it requires a sufficiently large number of simulations
or observations. Therefore, a PDF is usually constructed by using data from a
long time series, and in a technical term, is stationary (independent of time).
Unfortunately, such stationary PDFs (averaged over time) miss crucial
information about the dynamics/evolution of non-equilibrium processes
(e.g.\ tumour evolution). Similar difficulty often arises in theoretical
calculations due to analytical intractability, most previous work focusing on
the computation of stationary PDFs. However, as the advancement of experimental
techniques (e.g.\ single-cell technology) enables us to access variability in
tumour growth, different gene expression \cite{SHALEK14}, cellular processes,
etc., it is essential to predict time-dependent PDFs to improve our
understanding of fundamental mechanisms determining the evolution of different
distributions. Mathematically, in our growth models, a stochastic noise can
appear either multiplicatively or additively, representing different types of
internal or external noise.\\

In this paper, we calculate time-dependent PDFs in the stochastic logistic and
Gompertz models and explore similarities and disparities between the two models.
In particular, we demonstrate that stochastic noise can induce an interesting
transition from unimodal to bimodal PDFs in both models as the attractor around
the stable equilibrium point becomes leaky. For purely multiplicative noise, a
much weaker (leaky) attractor in the Gompertz model is shown to lead to a much
more significant (singular) growth of the population of a very small size.
We show that time-dependent PDFs can drastically differ from stationary PDFs
and that the variance is not a useful representation of the variability of a
bimodal PDF, highlighting the importance of time-dependent PDFs, as well as
other diagnostic quantities besides simply mean value and variance.
We also present the geometric methodology to understand a growth process from
the perspective of information change, and compare these two models in terms of
information change. The latter is quantified by the information length which
represents the total number of statistically different states that the system
undergoes in time (see Section IV). This information length provides us with a
useful system-independent method of understanding different stochastic
processes to analyse different experimental data. In particular, we show that
even though the time-evolution of the two models can be very different, they
become more similar when measured in units of the information length. We also
discuss implications of our results for bimodal variability in gene expression
observed in single-cell experiments (e.g.\ \cite{SHALEK14}) and persisters
(e.g.\ \cite{persister}).\\

The remainder of the paper is organised as follows. Section II introduces our
stochastic logistic and Gompertz models. Section III presents exact analytic
solutions for purely multiplicative noise, and shows that they develop
exponentially growing (and correspondingly narrowing) peaks, as well as a
singularity at the origin for the Gompertz model. Section IV reviews the
concept of information length, and applies it to the previous solutions.
Section V considers the case of both multiplicative and additive noise, and
shows how any non-zero additive noise regularises the previous solutions, and
ensures that the final evolution is always toward a non-singular stationary
distribution. Technical details of some of the derivations are also provided
in two appendices.\\

\section{Stochastic Logistic and Gompertz models}
To include stochastic noise in the logistic model, we consider the following
Langevin equation:
\begin{eqnarray}
\frac{dx}{dt} &=& (\gamma+\xi_1)x - \epsilon (\gamma +\xi_2)x^2+ \xi_3.
\label{eq1}
\end{eqnarray}
Here, as in Eq.\ (\ref{eq001}), $\gamma$ and $\gamma \epsilon$ are positive
constants, representing the constant part of the growth rate and nonlinear
damping coefficient. $\xi_1$ and $\xi_2$ are stochastic parts of these growth
rate and damping coefficients. $\xi_3$ is an additive noise modelling
fluctuations in the environment. We assume that $\xi_i$ has zero mean value
$\la \xi_i \ra = 0$, where angular brackets $\la \ra$ denote average over the
stochastic noise. Furthermore, for simplicity, we assume that $\xi_i$
($i=1,2,3$) are Gaussian noises with short correlation time and the following
correlation function
\begin{eqnarray}
\langle \xi_i(t) \xi_j (t') \rangle &&=D_{i,j} \delta(t-t'),
 \label{eq2} 
\end{eqnarray}
for $i=1,2,3$. $D_i$ represents the amplitude of the stochastic noise $\xi_i$,
while $D_{ij}$ represents the amplitude of the cross-correlation between
$\xi_i$ and $\xi_j$. The statistically independent noises $\xi_i$ and $\xi_j$
for $i \ne j$ are thus represented by $D_{ij} = 0$. Different cases of
$D_{ij}$ were studied in previous works, although they tend to be limited to
stationary PDFs. In particular, bimodal stationary PDFs were shown for
$D_{13} \ne 0$ and $\xi_2=0$ \cite{AI03}. The time-dependent PDFs were
shown in our previous work for a linear model (i.e. $\epsilon=\xi_2=\xi_3=0$)
\cite{KIM15} and for a nonlinear model with $\xi_1=\xi_3=0$
\cite{KIM15,HK16,KIM16}.\\

Similarly, as a generalisation of Eq.\ (\ref{eq003}), we consider the following
Langevin equation for the Gompertz model for a stochastic variable $y$:
\begin{eqnarray}
\frac{dy}{dt} &=& (\gamma + \xi_1)y - \epsilon (\gamma + \xi_2)y \ln{y} + \xi_3.
\label{eq3}
\end{eqnarray}
Compared with the stochastic logistic model in Eq.\ (\ref{eq2}), much less
analysis has been done on the stochastic Gompertz model. We show later some of
the challenges in obtaining solutions to Eq.\ (\ref{eq3}) and the corresponding
Fokker-Planck equation.

In the following, we consider two different cases of stochastic noises.
Specifically, we investigate the case of the same multiplicative noise
$\xi_1=\xi_2 = \xi$ for the growth rate and nonlinear damping with no
additive noise $\xi_3=0$ in Section III. We then include an independent
additive noise $\xi_3 \ne 0$ with $D_{12} = D_{13} = 0$ in this model in
Section V.

\section{Multiplicative noise only ($\xi_1=\xi_2 = \xi$ and $\xi_3=0$)}
In this section, we consider the case of the same multiplicative noise
$\xi_1=\xi_2 = \xi$ for the growth rate and nonlinear damping, with no
additive noise $\xi_3=0$. This case $\xi_1=\xi_2 = \xi$ is obtained in the
limit of the strong correlation between $\xi_1$ and $\xi_2$ with
$D_{11}=D_{22}=D_{12}\equiv D$. Therefore, Eq.\ (\ref{eq1}) and (\ref{eq3})
are reduced to
\begin{eqnarray}
\frac{dx}{dt} &=& (\gamma + \xi)x(1 - \epsilon x),
\label{eq1b}\\
\frac{dy}{dt} &=& (\gamma + \xi)y(1 - \epsilon \ln{y}).
\label{eq3b}
\end{eqnarray}
In the seminal work by Kimura \cite{kimura64}, Eq.\ (\ref{eq1b}) was used to
model a random drift (selection) in population genetics. The Fokker-Planck
equations corresponding to Eqs. (\ref{eq1b}) and (\ref{eq3b}) are as follows:
\begin{eqnarray}
{\partial_t p(x,t)}
&=& -{\partial_x} \left[\gamma x (1-\epsilon x)p(x,t) \right]
 + D \partial_x\Bigl[x(1-\epsilon x) \partial_x [x(1-\epsilon x) p(x,t)]\Bigr],
\label{eq2000}\\
{\partial_t p(y,t)}
&=& -{\partial_y} \left[\gamma y (1-\epsilon \ln{y})p(y,t) \right]
 +D \partial_y\Bigl[y(1-\epsilon \ln{y}) \partial_y
               [y(1-\epsilon \ln{y}) p(y,t)]\Bigr].
\label{eq2001}
\end{eqnarray}
In the following section we will derive exact solutions to these two sets
of Langevin and corresponding Fokker-Planck equations.

\subsection{Time-dependent PDFs}
We use the Stratonovich calculus \cite{Klebaner,Gardiner,WongZakai}, which
recovers the limit of a short correlated forcing from the finite correlated
forcing (e.g.\ \cite{WongZakai}), and show how to obtain the exact
solution for $x$ and $y$ and their time-dependent PDFs.

First, to solve Eq.\ (\ref{eq1b}), we divide it by $x(1-\epsilon x)$:
\begin{eqnarray}
\left[ \frac{1}{x} + \frac{\epsilon}{1-\epsilon x}\right] \frac{dx}{dt}
&=& \gamma + \xi.
\label{eq4}
\end{eqnarray}
By integrating Eq.\ (\ref{eq4}) over time, we obtain
\begin{eqnarray}
\ln{\left| \frac{x(1-\epsilon x_0)}{(1-\epsilon x)x_0} \right|}
 &=& \gamma t + \Gamma(t),
\end{eqnarray}
or, alternatively,
\begin{eqnarray}
x(t) &=& \frac{x_0}{\epsilon x_0 +(1-\epsilon x_0)
 \exp{(-\gamma t- \Gamma (t))}}\,,
\label{eq5}
\end{eqnarray}
where $\Gamma(t) = \int_0^t dt_1 \xi (t_1)$ is a Brownian motion. We note that
the probability distribution function (PDF) of $\Gamma$ is given by a Gaussian
distribution with the zero mean value and the inverse temperature $\beta$ as
\begin{eqnarray}
p(\Gamma,t) &=& \sqrt{\frac{\beta}{\pi}} \exp{\left [-\beta \Gamma^2\right]},
\label{eq500}
\end{eqnarray}
where $\beta = {1}/{(4Dt)}$ (e.g.\ \cite{Fokker}). $\beta$ is related
to the PDF's width or standard deviation $\sigma$ as $\beta={1}/{2\sigma^2}$.\\

The probability distribution function of $x$ is then obtained from
Eq.\ (\ref{eq4}) by using $p(x) = p(\Gamma)\left|\frac{d\Gamma}{dx}\right|$
and by expressing $\Gamma$ in terms of $x$ with the help of Eq.\ (\ref{eq5}):
\begin{equation}
p(x,t) = \sqrt{\frac{\beta}{\pi}} \frac{1}{|x(1-\epsilon x)|}
 \exp{\left [-\beta\left(\ln{ \left | \frac{x(1-\epsilon x_0)}{(1-\epsilon x)x_0}
 \right |}- \gamma t \right )^2\right]}
\equiv  \sqrt{\frac{\beta}{\pi}} \exp{[-\phi]},
\label{eq6}
\end{equation}
where $\phi$ is defined as
\begin{eqnarray}
\phi &=& \beta\left(\ln{ \left | \frac{x(1-\epsilon x_0)}{(1-\epsilon x)x_0}
 \right |}- \gamma t \right )^2 + \ln{|x(1-\epsilon x)|}.
\label{eq600}
\end{eqnarray}

Next, to solve  Eq.\ (\ref{eq3b}), let $z = \ln{y}$ to recast it as
\begin{equation}
\frac{dz}{dt} = (\xi + \gamma) (1 - \epsilon z).
\label{eq7}
\end{equation}
Again using the Stratonovich calculus \cite{Klebaner,Gardiner,WongZakai}, the
solution is found to be
\begin{equation}
\frac{1}{\epsilon} \ln{ \left| \frac{1- \epsilon z}{1- \epsilon z_0}\right| }
 = - \gamma t + \Gamma(t),
\label{eq8}
\end{equation}
\begin{equation}
y =  \exp \left[ \frac{1}{\epsilon}-\left( \frac{1}{\epsilon}
 - \ln{y_0} \right) e^{\epsilon [-\gamma t + \Gamma(t)]} \right],
\label{eq9}
\end{equation}
where $z_0 = z(t=0)$, $y_0 = y(t=0)$, and $\Gamma (t) = \int_0^t dt_1\,
\xi(t_1)$ is the Brownian motion.\\

To obtain the PDF $p(y,t)$, we first use the Gaussian PDF of $\Gamma(t)$ given
in Eq.\ (\ref{eq500}) and the conservation of the probability $p(z,t) = 
p(\Gamma) \left| \frac{d\Gamma}{dz} \right|$ together with
$\left| \frac{ d\Gamma}{dz}\right| = \frac{1}{|1-\epsilon z|}$ to obtain 
\begin{eqnarray}
p(z,t) &=& \sqrt{\frac{\beta}{\pi}}
\frac{1}{|1-\epsilon z|} \exp{\left [-\beta\left(\frac{1}{\epsilon} \ln{\left|
 \frac{1-\epsilon z}{1-\epsilon z_0} \right| }+ \gamma t \right )^2\right]}. 
\label{eq10}
\end{eqnarray}
We then use the conservation of the probability again as $p(y,t) = p(z,t)
\left|\frac{dz}{dy}\right|$ and $\frac{dz}{dy} = \frac{1}{y}$ (recall
$z=\ln{y}$) to obtain
\begin{equation}
p(y,t) = \sqrt{\frac{\beta}{\pi}}
\frac{1}{|y(1-\epsilon \ln{y})|} \exp{\left [-\beta\left(\frac{1}{\epsilon}
\ln{\left| \frac{1-\epsilon \ln{y}}{1-\epsilon \ln{y_0}} \right| }
+ \gamma t \right )^2\right]}
\equiv  \sqrt{\frac{\beta}{\pi}} \exp{[-\psi]},
\label{eq112}
\end{equation}
where $\psi$ is defined as
\begin{eqnarray}
\psi &=& \beta\left(\frac{1}{\epsilon} \ln{\left| \frac{1-\epsilon
 \ln{y}}{1-\epsilon \ln{y_0}} \right| }+ \gamma t \right )^2
 + \ln{|y(1-\epsilon \ln{y})|}.
\label{eq113}
\end{eqnarray}

To summarize, Eqs.\ (\ref{eq6}) and (\ref{eq112}) are derived as the PDFs of
the Langevin equations (\ref{eq1b}) and (\ref{eq3b}), and are correspondingly also
exact solutions of the Fokker-Planck equations (\ref{eq2000}) and (\ref{eq2001})
respectively (as can also be verified by direct substitution back into these
equations). Both solutions represent the evolution of $\delta$-functions initially
located at $x_0$ and $y_0$, respectively.

Figure 1 shows examples of the subsequent evolution. Here and throughout the
remainder of this paper we fix $\gamma=1$ for both processes, and $\epsilon=1/e$
for logistic and $\epsilon=1$ for Gompertz to ensure that both models have the
same stable equilibrium point $e$ in the absence of the stochastic noise. 
Our expectation therefore is that
solutions should evolve toward the stable equilibrium at $e$. As seen in Figure 1,
for small $D$ this is indeed the case; the $\delta$-functions initially located at
$x_0=y_0=0.8$ move monotonically toward $e$. In the intermediate stages of the
evolution they also broaden considerably, but once they approach $e$ they become
increasingly narrow again.

In contrast, for larger $D$ the peaks hardly move at all, but instead become so
broad that they fill essentially the entire interval $(0,e)$. Once they sense
the presence of the equilibrium at 0, they also begin to form peaks there, even
though 0 is an {\it unstable} point of the original equations (\ref{eq001}) and
(\ref{eq003}). Indeed, for the Gompertz model the peak at the origin is clearly
far stronger than the peak at $e$. For the logistic model the peaks are more
comparable, and Figure 1 is inconclusive regarding which one is ultimately
dominant.

\begin{figure}
\centering
\includegraphics[scale=0.9]{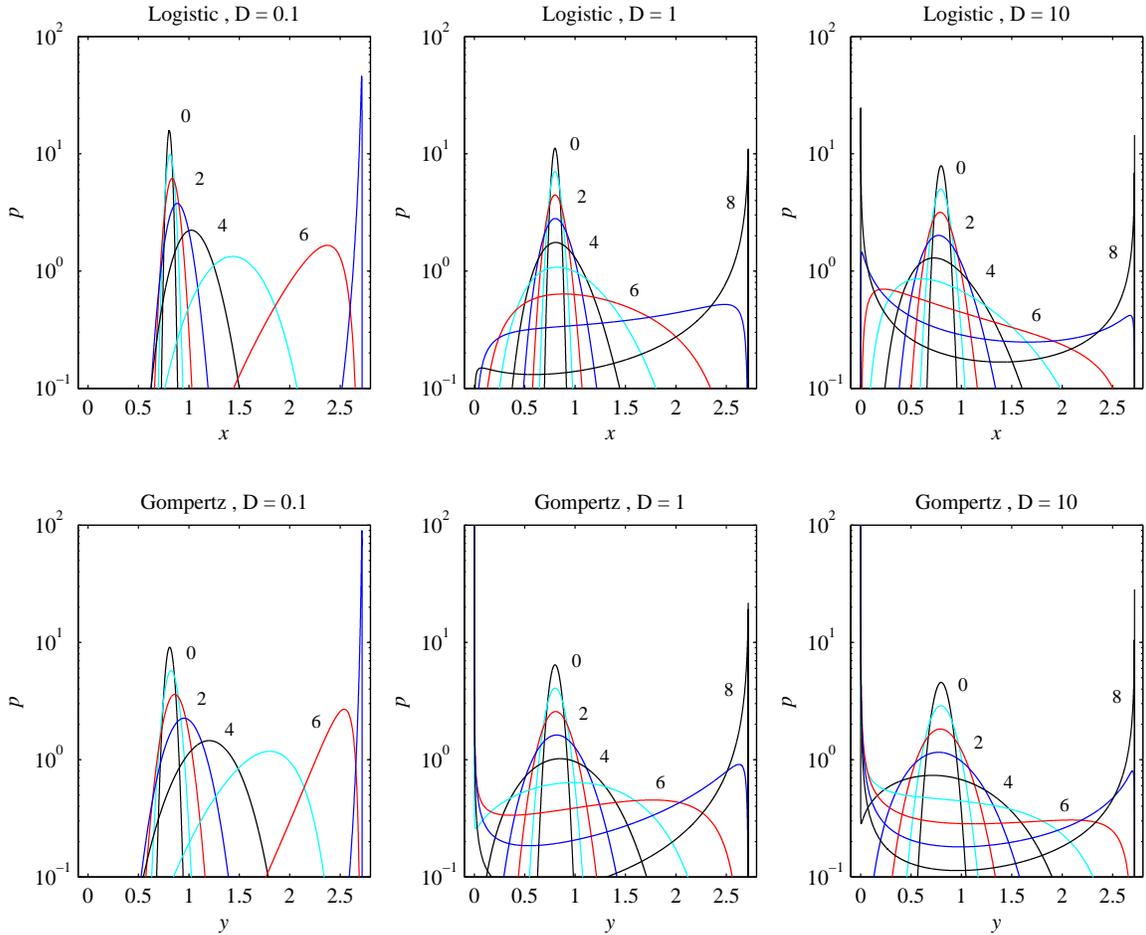}
\caption{Illustrations of the analytic expressions (\ref{eq6}) for the logistic
model in the top row, and (\ref{eq112}) for the Gompertz model in the bottom row.
From left to right $D=0.1$, $1$ and $10$ as indicated. The $\delta$-functions
are initially located at $x_0=y_0=0.8$. The numbers beside individual curves
indicate time: For $D=0.1$ these are given by $t=0.01\cdot2.5^n,\ n=0-7$ (so
$t=0.01-6.1$), for $D=1$, $t=0.002\cdot2.5^n,\ n=0-8$ (so $t=0.002-3$), and for
$D=10$, $t=0.0004\cdot2.5^n,\ n=0-8$ (so $t=0.0004-0.61$).}
\vskip 1cm
\label{Fig1}
\end{figure}

\subsection{Behaviour of Peaks near $x=y=e$ and $x=y=0$}
We clearly need to better understand the behaviour of these solutions
(\ref{eq6}) and (\ref{eq112}), and especially this somewhat counter-intuitive
behaviour that for sufficiently large $D$ they become bimodal, yielding peaks
not only at the (expected) stable point $e$, but also at the unstable point $0$.

We begin with the logistic model (\ref{eq6}), and investigate its behaviour
near $e$. If we let $x=e-\delta x$ and expand the various terms in $\phi$ in
Eq.\ (\ref{eq600}) to first order in $\delta x$ we obtain
\begin{eqnarray}
\phi &\approx & \beta\left(\ln{ \left |
 \frac{(1-\epsilon x_0)}{ \epsilon^{2}\, \delta x\, x_0}
 \right |}- \gamma t \right )^2 + \ln{(\delta x)}.
\label{eq200}
\end{eqnarray}
The location of the PDF peak will occur at that $x$ where $\partial_{x} p(x,t)
= 0 = \partial_{x} \phi$, so differentiating (\ref{eq200}) gives us the motion
of the peak position as
\begin{eqnarray}
&& \frac{\partial \phi}{\partial x} \approx -2 \beta\left(\ln{ \left |
 \frac{(1-\epsilon x_0)}{ \epsilon^{2}\, \delta x\, x_0} \right |}-
 \gamma t \right ) \frac{1}{\delta x} + \frac{1}{\delta x} = 0,
\end{eqnarray}
which yields
\begin{eqnarray}
\delta x_s &\approx & \frac{1 - \epsilon x_{0}}{\epsilon^{2} x_{0}}
 e^{-\left (\gamma t + \frac{1}{2 \beta} \right) }=
 \frac{e ( e- x_{0})}{ x_{0}} e^{-(2D+1)t},
\label{eq2002}
\end{eqnarray}
where we have again used $\beta=1/4Dt$, $\gamma=1$ and $\epsilon=1/e$ for the
logistic model. With the position of the peak known, it is then straightforward
to evaluate the PDF at that position to obtain
\begin{equation}
p(e - \delta x_s,t)=
\sqrt{\frac{1}{ 4 \pi D t}} \,\frac{x_0}{e( e- x_0)}\,e^{(D+1)t}.
\label{eq2004}
\end{equation}

To consider the logistic model near the origin, we follow much the same
procedure, except that we expand in $x$ itself rather than in $\delta x$. The
results are
\begin{eqnarray}
\phi &\approx & \beta\left(\ln{ \left |
 \frac{x (1-\epsilon x_0)}{x_{0}} \right |}- \gamma t \right )^2 + \ln{x},
\label{eq2003}
\end{eqnarray}
so again setting the derivative to zero yields
\begin{eqnarray}
\frac{\partial \phi}{\partial x} &\approx & 2 \beta\left(\ln{ \left |
 \frac{x (1-\epsilon x_0)}{x_{0}} \right |}- \gamma t \right )
 \frac{1}{x} + \frac{1}{x} = 0,
\label{eq204}
\end{eqnarray}
\begin{eqnarray}
x_u &\approx & \frac{x_{0}}{1-\epsilon x_{0}} \,e^{\left (\gamma t
 - \frac{1}{2 \beta} \right) }= \frac{e x_{0}}{ e- x_{0}} e^{-(2D-1)t},
\label{eq205}
\end{eqnarray}
and the amplitude is finally given by
\begin{equation}
p(x_u,t)=\sqrt{\frac{1}{ 4 \pi D t}} \,\frac{e- x_0}{e x_0}\,e^{(D-1)t}.
\label{eq2009}
\end{equation}

If we then compare the peak amplitudes (\ref{eq2004}) and (\ref{eq2009}), we
see that the peak at the stable point $x=e$ will always dominate for sufficiently
large $t$, since it has the larger exponential growth rate. Nevertheless, for
$D>1$ the peak at the unstable point $x=0$ will also grow, indicating the
transition from a unimodal to a bimodal PDF.  That is, a sufficiently strong
multiplicative noise promotes a growth of the population ($x$) of small size
around the unstable equilibrium point.

Finally, if we compare the multiplicative factors, $x_0/e(e-x_0)$ for the peak
at $e$ versus $(e-x_0)/ex_0$ for the peak at $0$, we see that even these are
intuitively understandable: an $x_0$ close to $e$ means the $e$-peak has a
larger factor, whereas an $x_0$ close to $0$ means the $0$-peak has a larger
factor. Figure 2 shows how these asymptotic formulas compare with the exact
expression, and shows that even quite modest $t$ values already yield excellent
agreement. Other choices of $D$ and/or $x_0$ yielded similarly good agreement.
We conclude therefore that the behaviour of Eq.\ (\ref{eq6}) is fully
understood.

\begin{figure}
\centering
\includegraphics[scale=0.9]{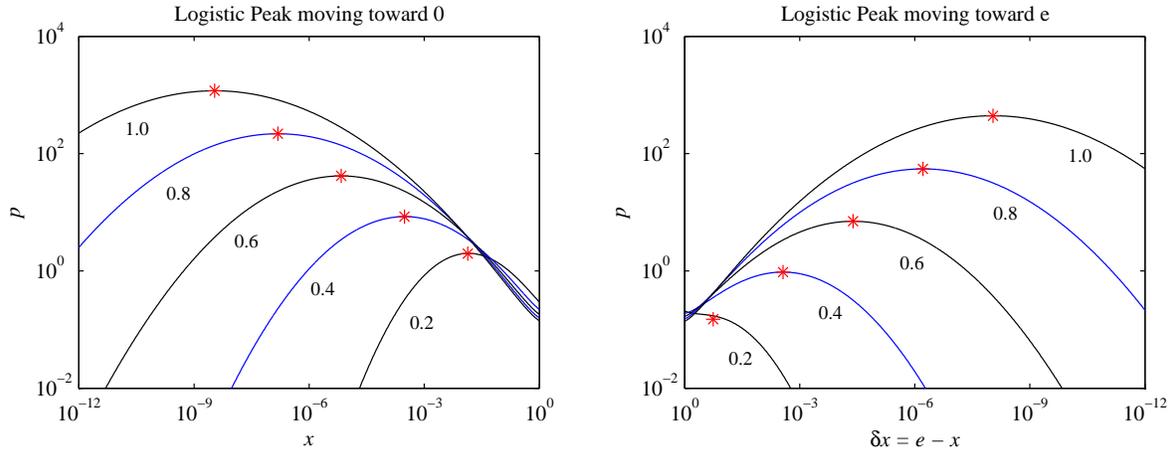}
\caption{The five curves in each panel correspond to the exact logistic solution
(\ref{eq6}), with $D=10$ and $x_0=0.5$, at the times indicated by the numbers
beside each curve. The asterisks denote the asymptotic formulas, (\ref{eq205})
and (\ref{eq2009}) for the approach to $0$ in the left panel, and (\ref{eq2002})
and (\ref{eq2004}) for the approach to $e$ in the right panel. Note how each
asterisk is indeed in near perfect agreement with the top of its corresponding
curve.}
\vskip 1cm
\label{Fig2}
\end{figure}

The Gompertz peak moving toward $e$ can be analysed in exactly the same way. If
we let $y=e-\delta y$ and expand $\psi$ in Eq.\ (\ref{eq113}) to first order in
$\delta y$, the final results are
\begin{equation}
\delta y_s \approx e(1 - \ln{y_{0}}) e^{- \epsilon \left
 (\gamma t + \frac{1}{2 \beta }\right) }= e(1 - \ln{y_{0}}) e^{-(2D+1)t},
\label{eq302}
\end{equation}
\begin{equation}
p(e - \delta y_s,t)=
 \sqrt{\frac{1}{ 4 \pi D t}}\,\frac{1}{e(1-\ln{y_0})}\,e^{(D+1)t}.
\label{eq3020}
\end{equation}
It is interesting to note that the exponential factors $e^{-(2D+1)t}$ and
$e^{(D+1)t}$ are both exactly the same as for the logistic peak moving toward
$e$. The stable equilibrium point is evidently insensitive to the precise form
of the nonlinear terms in the original Langevin equations. The agreement
between the exact solution (\ref{eq112}) and these asymptotic formulas
(\ref{eq302}) and (\ref{eq3020}) is also much the same as for the logistic
peaks in Figure 2.

In contrast, the behaviour of the Gompertz solution (\ref{eq112}) near the
origin is radically different. Instead of a peak moving toward the origin,
there is a powerful singularity that starts at the origin and moves inward.
To see this, we begin by evaluating $\frac{dp}{dy}$, setting it equal to 0,
and solving for $t$. The result is
\begin{equation}
t=-{\ln\left(\frac{1-\ln y}{1-\ln y_0}\right)}\Big/{(1+2D\ln y)}.
\label{tformula}
\end{equation}
If this equation could be inverted for $y$ as a function of $t$, it would give
an expression for the movement of any peaks (or troughs). However, even without
being able to invert for $y(t)$, by simply graphing $t(y)$ we can still track
the peaks and troughs.

Figure 3 shows representative examples, for $D=1$, and three $y_0$ values.
Starting with the simplest case, for $y_0=0.7$ the original $\delta$-function
moves monotonically outward, eventually approaching $e$ in precise agreement
with Eq.\ (\ref{eq302}). Additionally, there is a trough that originates from
$y\to0$, and eventually asymptotes to $y=e^{-1/2D}$ (where the denominator of
Eq.\ (\ref{tformula}) is zero). The first panel in the bottom row of Figure 3
shows the corresponding PDFs, which indeed exhibit peaks and troughs exactly as
predicted by the first panel in the top row.

Turning next to $y_0=0.5$, the two solution tracks of Eq.\ (\ref{tformula}) now
connect differently. The initial $\delta$-function peak combines with the
trough coming from $y\to0$, destroying both. There is then an intermediate
time $0.195<t<0.525$ (for these particular $D$ and $y_0$ values) where there
are neither peaks nor troughs, and $p(y,t)$ is monotonic. Finally, for $t>0.525$
a new solution track emerges, with the peak again approaching $e$, and the
trough approaching $e^{-1/2D}$. The third panel in the bottom row of Figure 3
again shows the corresponding PDFs; note especially the monotonic behaviour at
the intermediate time.

Finally, the middle panels in Figure 3 show the transition point from one
regime to the other. For $y_0>e^{-1/2D}$ the solution tracks are like the
$y_0=0.7$ case, for $y_0<e^{-1/2D}$ they are like the $y_0=0.5$ case, and for
$y_0=e^{-1/2D}$ ($\approx0.61$ for $D=1$) they are as indicated in the
middle panel.

\begin{figure}
\centering
\includegraphics[scale=0.9]{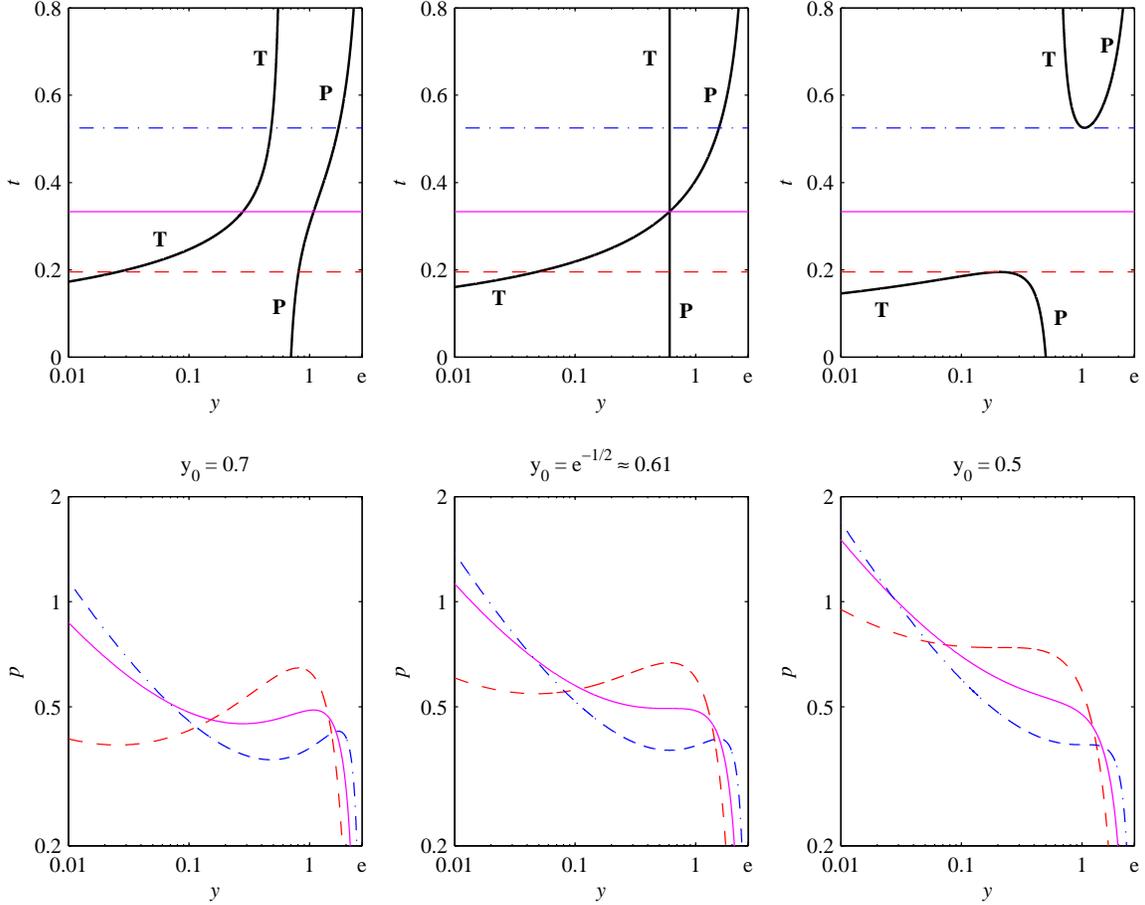}
\caption{The thick black lines in the top row plot Eq.\ (\ref{tformula}),
showing the locations in time of any peaks or troughs, indicated by the
letters {\bf P} and {\bf T}. From left to right $y_0=0.7$, 0.61 and 0.5, and
$D=1$ for all three. The thin horizontal lines are at $t=0.195$, 0.333 and
0.525, with the corresponding PDFs shown in the bottom row. Note how the
peaks and troughs (or lack thereof) agree with the predictions in the top row.
Note finally how the $t=0.525$ PDFs especially are strongly suggestive of a
$p\sim y^a$ scaling, with $a<0$, indicating a singularity at the origin.}
\vskip 1cm
\label{Fig3}
\end{figure}

For large $t$ we see therefore that the behaviour is always the same: there
is a peak approaching $e$ as described by Eq.\ (\ref{eq302}), and there is a
trough approaching $e^{-1/2D}$ from either the left or the right, depending
on whether $y_0$ is less than or greater than this value. The final point to
understand then is the trough emerging from the $y\to0$ regime, and what
$p(y,t)$ ultimately looks like in this regime. We begin by noting that
because Eq.\ (\ref{tformula}) has $\displaystyle\lim_{y\to0}t=0$, whatever
singular behaviour emerges from this region happens {\it instantaneously},
as soon as $t>0$.

To see what $p(y,t)$ looks like for $y\to0$, we note that the bottom row of
Figure 3 suggests a scaling of the form $p\sim y^a$, with $a<0$. This in turn
suggests examining the quantity $\frac{y}{p}\,\frac{dp}{dy}$; if $p$ were
exactly of the form $y^a$, then $a$ would equal exactly this quantity.
From (\ref{eq112}) we obtain
\begin{equation}
\alpha=\frac{y}{p}\,\frac{dp}{dy}
=-\frac{\ln\left(\frac{1-\ln y}{1-\ln y_0}\right) + (1+2D\ln y)t}
       {(1-\ln y)2Dt}.
\label{alphaformula}
\end{equation}
To interpret this result, we first note that $\alpha$ depends on $y$,
indicating that $p$ is not exactly of the form $y^\alpha$. However, because
$y$ enters into $\alpha$ only as $\ln y$, the dependence is very weak, and
locally $p$ is closely approximated by $y^\alpha$. Figure 4 shows $\alpha$ as
a function of $t$, for $D=0.1$, 1 and 10, and $y=10^{-10}$, $10^{-20}$ and
$10^{-30}$. For all combinations we find that $\alpha$ becomes negative once
$t>O(D^{-1})$. Indeed, if we are interested in the true $y\to0$ limit, we find
$\alpha\to-1$ for all $t>0$. So again, the singularity at the origin starts
instantaneously, and propagates outward to ever larger $y$ as $t$ increases,
with
\begin{equation}
\lim_{t\to\infty}\alpha=\frac{1/2D + \ln y}{1-\ln y}.
\end{equation}
Note how this quantity varies between $-1$ for $y\to0$ and $0$ for
$y=e^{-1/2D}$, where it coincides with the previously understood behaviour
of the trough at that location. For all $y$ to the left of that trough
therefore $p$ is monotonically increasing, and indeed approaching ever
closer to a $y^{-1}$ singularity at the origin.  (Because $\alpha$ never
reaches $-1$ for any non-zero $y$ though, the integral $\int_0^\infty p\,dy$
not only converges, but always remains 1, as required by conservation of
total probability.)

\begin{figure}
\centering
\includegraphics[scale=0.9]{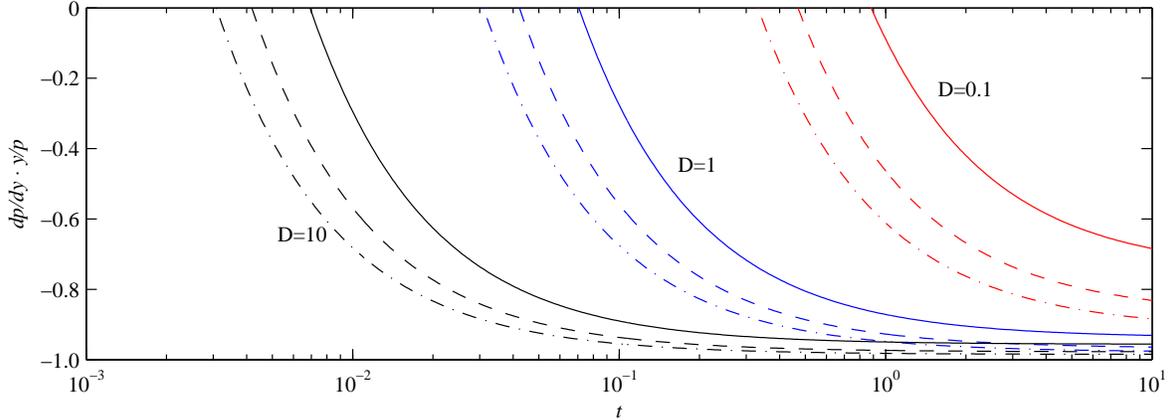}
\caption{The quantity $\alpha=\frac{y}{p}\,\frac{dp}{dy}$ in
Eq.\ (\ref{alphaformula}) as a function of time, for $D=0.1$, 1 and 10 as
indicated, and $y=10^{-10}$ (solid), $10^{-20}$ (dashed), and $10^{-30}$
(dash-dotted). Note in particular how $\alpha$ becomes negative for all
combinations of parameters, and how this happens earliest for $y\to0$.
The initial peak is fixed at $y_0=1$; other values yield qualitatively
the same behaviour.}
\vskip 1cm
\label{Fig4}
\end{figure}

\section{Information length}
In this section, we show how to utilise PDFs to understand information change
in general, and then apply these concepts to our logistic and Gompertz PDFs.
First, for any given two PDFs, we quantify the difference between them by
assigning an appropriate metric to probability such that the distance increases
with the difference between the two PDFs. This metric provides an interesting
link between stochastic processes and associated geometric structure. A key
characteristic of non-equilibrium processes is however in its variability in
time (or space), reflected in the temporal change in PDFs, time-varying
PDFs implying the change in information content in the system. In this case, we
define an infinitesimal distance at any time by comparing two PDFs at times
infinitesimally apart and sum these distances in time. The total distance along
the trajectory of the system quantifies the total number of different states
that the system undergoes in time, and is called the information length
\cite{HK16,KIM16,NK14,NK15,KH17}.\\

To show this explicitly, we consider a stochastic variable $z$ and suppose
that we can compute its time-dependent PDFs $p(z,t)$ either analytically or
numerically in the case where its governing equation is known, or otherwise
construct $p(z,t)$ from experimental/observational data. Defining the
information length involves two steps \cite{HK16,KIM16,NK14,NK15,KH17,HK17}:
First we need to compute the dynamic time unit $\tau(t)$, which is the
characteristic timescale over which $p(z,t)$ temporally changes {\it on
average} at time $t$. Second, we need to compute the total elapsed time in
units of this $\tau(t)$. As done in \cite{HK16,KIM16,NK14,NK15,KH17,HK17},
we compute $\tau$ by utilising the following second moment ${\cal E}$:
\begin{eqnarray}
{\cal{E}} \equiv \frac{1}{[\tau (t)]^2} & = &\int dz \frac {1} {p(z,t)}
 \left [\frac {\partial p(z,t)} {\partial t} \right]^2.\label{eq01}
\end{eqnarray}
We note that ${\cal E}$ is the root-mean-square fluctuating energy for a
Gaussian PDF (see \cite{KIM16}). As defined in Eq.\ (\ref{eq01}), $\tau$ has
dimensions of time, and quantifies the correlation time over which $p(z,t)$
changes, thereby serving as the time unit in statistical space. Alternatively,
$1/\tau$ quantifies the (average) rate of change of information in time. We
recall that $\tau(t)$ in Eq.\ (\ref{eq01}) is related to the second derivative
of the relative entropy (or Kullback-Leibler divergence) (see Appendix A and
\cite{HK16}).\\

The total accumulated change in information between the initial and final
times, $0$ and $t$ respectively, is defined by measuring the total elapsed
time in units of $\tau$ as:
\begin{equation}
{\cal{L}} (t) =\int_0^{t} {dt_1}\sqrt{{\cal E} (t_1)} = \int_0^{t}
 \frac{dt_1}{\tau(t_1)} = \int_0^{t} dt_1 \sqrt{\int dz \frac {1} {p(z,t_1)}
 \left [\frac {\partial p(z,t_1)} {\partial t_1} \right]^2}.
\label{eq02}
\end{equation}
Eq.\ (\ref{eq02}) provides the total number of different states that a system
passes through from the initial state with the PDF $p(z,0)$ at time $0$ to
the final state with the PDF $p(z,t)$ at time $t$, establishing a distance
between the initial and final PDFs in the statistical space. For example, in
equilibrium where $\frac{\partial p}{\partial t}=0$, ${\cal E}=0$ and hence
$\tau(t_1)\to\infty$ for all time $t_1$. Measuring $dt_{1}$ in units of this
infinite $\tau(t_1)$ at any $t_{1}$, $dt_1/\tau(t_1) = 0$ in Eq.\ (\ref{eq02}),
and thus $\int_0^t dt_1/\tau(t_1)=0$. This can be viewed as asserting that in
statistical space there is no flow of time in equilibrium. In the opposite
limit, large $\cal E$ corresponds to small $\tau$, meaning that information
changes very quickly in dimensional time. See Appendix A for the
interpretation of ${\cal L}$ from the perspective of the infinitesimal
relative entropy.\\

Our information length is based on Fisher information (c.f.~\cite{fisher}) and
is a generalisation of statistical distance \cite{WOOTTERS81}, where the
distance is set by the number of distinguishable states between two PDFs.
While the latter was heavily used in equilibrium or near-equilibrium of
classical and quantum systems (e.g.\ \cite{RUPPEINER79,Salamon02}), our
recent work \cite{HK16,KIM16,NK14,NK15,KH17,HK17} adapted this concept to a
non-equilibrium system to elucidate geometric structure of non-equilibrium
processes. Specifically, \cite{KH17} mapped out the attractor structure
${\cal L}_{\infty}$ vs $z_{0}$ for linear and cubic processes and showed that
a linear damping preserves a linear geometry ${\cal L}_{\infty} \propto z_{0}$
whereas a nonlinear damping gives rise to a power-law scaling
${\cal L}_{\infty} \propto z_{0}^n$ ($n \sim 1.5-1.9$) of the attractor
structure. \cite{KIM16} found interesting geodesic solutions in a
non-autonomous Ornstein-Uhlenbeck (O-U) process \cite{Fokker} by modulating model parameters
and by including time-dependent external deterministic killing term. Notably,
the modulation of the model parameters and the killing term were
periodic/oscillatory. \cite{KH17,HK17} reported the asymmetry in ${\cal L}$
in order-to-disorder versus disorder-to-order transitions.\\

To now apply these general ideas to our particular models here, we begin with
the analytical derivation of ${\cal E}$ and ${\cal L}$ for the Gompertz model
(\ref{eq112}). To derive $\cal E$ defined in Eq.\ (\ref{eq01}), we let
$z=\ln{y}$ and use Eq.\ (\ref{eq10}) to obtain:
\begin{equation}
\partial_t{p(z,t)} = \left \{ \dot{\beta} \left[ \frac{1}{2 \beta}
 - \Gamma (t) ^2 \right] -2 \gamma \beta \Gamma(t) \right \} p(z,t),
\label{eqpdot}
\end{equation}
where $\Gamma (t) = \int_0^t dt_1\,\xi(t_1) = \frac{1}{\epsilon}
 \ln{\frac{1-\epsilon z_0}{1-\epsilon z}} -\gamma t$ is the Brownian motion.
Using Eq.\ (\ref{eqpdot}), we then obtain
\begin{eqnarray}
{\cal E} &=& \int \frac{(\partial_t p(y,t) )^2}{p(y,t)} dy
= \int \frac{(\partial_t p(z,t))^2}{p(z,t)} dz 
\nonumber\\
&=& \dot{\beta}^2 \left \langle \left( \frac{1}{2 \beta}
 - \Gamma ^2 \right)^2 \right \rangle - 4\gamma \beta \left \langle \Gamma
 \left( \frac{1}{2 \beta} - \Gamma ^2 \right ) ^2 \right \rangle
 + 4 \gamma ^2 \beta ^2 \left \langle \Gamma ^2 \right \rangle 
\nonumber\\
&=& \dot{\beta}^2 \left [ \frac{1}{4 \beta ^2} - \frac{1}{2 \beta ^2}
 + \frac{3}{4 \beta ^2} \right ] + 2 \gamma ^2 \beta 
= \frac{\dot{\beta}^2 }{2 \beta^2}+ 2 \gamma^2 \beta
= \frac{\dot{\beta}^2 }{2 \beta^2}+ 2\beta\left\langle\dot{z}\right\rangle ^2.
\label{eq12}
\end{eqnarray}
Here we used ${\langle \dot{z} \rangle}=\gamma$, $\langle \Gamma \rangle =
 \langle \Gamma^3 \rangle = 0$, $\langle \Gamma^2 \rangle = 1/2 \beta$, and
$\langle \Gamma ^4 \rangle = 3 \langle \Gamma^2 \rangle$. 
It is interesting to note that in terms of the inverse temperature $\beta$ and
the mean value $\langle z \rangle$, $\cal E$ in Eq.\ (\ref{eq12}) takes the
same form as for $\cal E$ in the Ornstein-Uhlenbeck process
\cite{Fokker}. Using $\beta = {1}/{4Dt}$ for our model, we
simplify Eq.\ (\ref{eq12}) as follows:
\begin{equation}
{\cal E} = \frac{1}{2\beta ^2} \left(\frac{\beta}{t}\right) ^2
 +2 \gamma \beta = \frac{1}{2 t^2} +2 \gamma \beta
  = \frac{1}{2 t^2} \left(1 + \frac{\gamma}{D} t \right).
\label{eq13}
\end{equation}

By using the definition of ${\cal L}(t_i,t_f) = \int_{t_i}^{t_f} dt_1
\sqrt{{\cal E}(t_1)}$ given in Eq.\ (\ref{eq02}) and using Eq.\ (\ref{eq13}),
we obtain $\cal L$ in the following form:
\begin{equation}
{\cal L}(t)=\int_{t_i}^{t_f} dt \frac{1}{\sqrt{2} t}
 \sqrt{1+ \overline{\gamma} t},
\label{eq130}
\end{equation}
where $\overline{\gamma}={\gamma}/{D}$. We let
$T=\sqrt{1 +\overline{\gamma}t}$ and $T_i = T(t=t_i)$ and $T_f = T(t=t_f)$
and use the change of variables to evaluate Eq.\ (\ref{eq130}) as follows:
\begin{eqnarray}
{\cal L} &=& \frac{2}{\gamma}
\int _{T_i}^{T_f} dT \ T \frac{\gamma}{\sqrt{2}}\frac{T}{T^2 - 1}
 = \sqrt{2} \int _{T_i}^{T_f} dT \left [ 1 + \frac{1}{T^2 -1} \right ]
\nonumber\\
&=&\sqrt{2} \left[ T + \frac{1}{2} \ln{\frac{T-1}{T+1}} \right ] _{T_i}^{T_f} 
   = \sqrt{2} \left \{ T_f - T_i + \frac{1}{2}
 \ln{ \left[ \frac{T_f - 1}{T_f + 1} \frac{T_i +1}{T_i - 1} \right ]} \right \},
\label{eq14}
\end{eqnarray}
where $T_i=\sqrt{1+{\gamma t_i}/{D}}$ and $T_f=\sqrt{1+{\gamma t_f}/{D}}$.\\

Next, we show that {\it identical} results are obtained for the stochastic
logistic model by following similar analysis as above. First, from
Eq.\ (\ref{eq6}), we obtain
\begin{equation}
\partial_t {p(x,t)} = \left \{ \dot{\beta} \left[ \frac{1}{2 \beta}
 -{\Gamma} ^2 \right] -2 \gamma \beta \Gamma \right \} p(x,t),
\end{equation}
where ${\Gamma}(t) = \int_0^t dt_1 \, \xi(t_1) =
\ln{\frac{(1-\epsilon x_0)x}{(1-\epsilon x)x_0}} -\gamma t$ 
is the Brownian motion. Thus, 
\begin{eqnarray}
{\cal E} &=& \int \frac{(\partial_t{p(x))}^2}{p(x)} dx 
= \dot{\beta}^2 \left \langle \left( \frac{1}{2 \beta} - {\Gamma} ^2 \right)^2
 \right \rangle - 4\gamma \beta \left \langle {\Gamma} \left( \frac{1}{2 \beta}
 - {\Gamma} ^2 \right ) ^2 \right \rangle + 4 \gamma ^2 \beta ^2
 \left \langle{\Gamma} ^2 \right \rangle 
\nonumber\\
&=& \frac{\dot{\beta}^2 }{2 \beta^2}+ 2 \gamma^2 \beta
 = \frac{1}{2 t^2} \left(1 + \frac{\gamma}{D} t \right),
\label{eq120}
\end{eqnarray}
where again $\beta = {1}/{4Dt}$, which is the same as Eq.\ (\ref{eq12}).
Eq.\ (\ref{eq120}) thus leads to the same information length ${\cal L}$ as in
Eq.\ (\ref{eq14}).\\

It is quite extraordinary that two processes as different as the logistic and
Gompertz models should nevertheless have exactly the same functions
${\cal E}(t)$ and ${\cal L}(t)$. This is due to the fact that Eqs. (\ref{eq1b}) and
(\ref{eq3b}) can be mapped into a similar Gaussian process by a suitable change
of variables. Another interesting -- and {\it a priori} not
obvious -- point is that the initial positions $x_0$ and $y_0$ do not enter
into these expressions. All peaks starting anywhere for either model have the
same ${\cal L}(t)$, with $D$ being the only remaining parameter. Figure 5
shows the results. For $t$ up to $O(1)$, $\cal L$ is independent of $D$, and
scales as $\ln t$.  (This also means that $t_i=0$ should not be used in
Eq.\ (\ref{eq14}); the value used here is $t_i=10^{-5}$.) For sufficiently
large $t$ we recover the $\sqrt{t/D}$, in agreement of the scaling predicted
 in (\ref{eq14}).
This independence of ${\cal L}$ on $x_0$ and $y_0$ can be traced back to the fact
that the movement of the PDF peak in the transformed Gaussian process is a drift, 
which is independent of the position,
in a sharp contrast to the movement of PDF peak in the O-U process 
caused by the position-dependent frictional force.
These results thus reveal the merit of using the information length
in unifying different stochastic processes. \\

Finally, if we return briefly to Figure 1, two points stand out: First, the
entire evolution shown in Figure 1 occurs in this $t\leq O(1)$ regime. The
$\sqrt{t/D}$ scaling obtained for $t>O(1)$ therefore only applies to peaks
that are already so narrow that they would be unlikely to be relevant to
real-world data. Second, we recall that the curves shown in Figure 1 are at
times $t=t_0\cdot2.5^n$, $n=0,1,\ldots$. The ${\cal L}\approx0.71\ln(t/t_i)$
scaling for small $t$ then implies that the information length between
successive PDFs is roughly constant, around $0.71\ln{2.5}=0.65$. 
In fact, the time $t=t_0\cdot2.5^n$, $n=0,1,\ldots$ in Figure 1 was chosen
precisely to show the evolution of a PDF with the equal increment of ${\cal L}$,
so that the two PDFs at the two subsequent times have the same change in
information. Figure 1 shown at equal increments of time would look very
different, and much less informative as some of the PDFs would look very
similar while others would look drastically different. This highlights another
advantage of using the information length in understanding information flow
and true dynamical change in non-equilibrium processes. 

\begin{figure}
\centering
\includegraphics[scale=0.9]{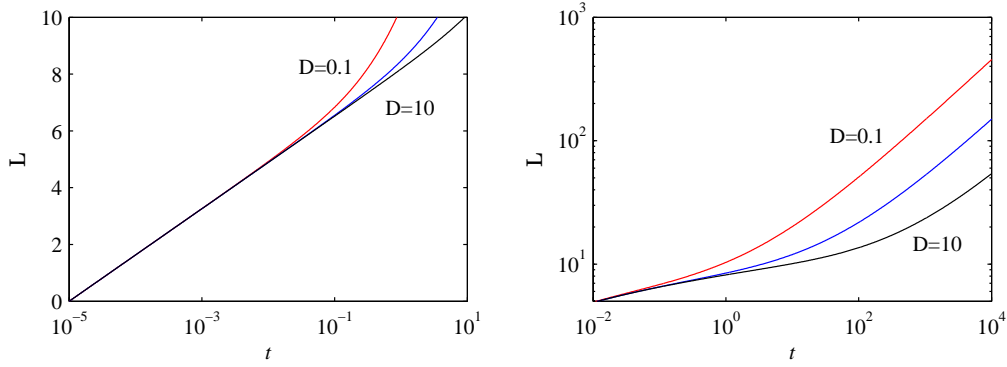}
\caption{${\cal L}(t)$ according to Eq.\ (\ref{eq14}), for $D=0.1$, 1 and 10
as indicated. The left panel shows the $\ln t$ small-time behaviour; the right
panel shows the $\sqrt{t/D}$ large-time behaviour.}
\vskip 1cm
\label{Fig5}
\end{figure}

\section{Multiplicative and additive noise
($\xi_1 = \xi_2 = \xi$ and $\xi_3\neq 0$)}
We saw in section III that the models with only internal but no external noise
never form stationary PDFs, but instead form ever sharper peaks, eventually
becoming so narrow that they are hardly relevant to most real situations.
In this section we therefore extend these models to include an additive noise
$\xi_3\equiv\eta$, which can represent either external noise or random mutation
rates (e.g. \cite{ZHANG10}). In this case, Eqs. (\ref{eq1}) and (\ref{eq3})
take the form:
\begin{eqnarray}
\frac{dx}{dt}&=& (\gamma + \xi) x (1 - \epsilon x) + \eta,
\label{eq1c}\\
\frac{dy}{dt}&=& (\gamma + \xi) y (1 - \epsilon \ln{y}) + \eta,
\label{eq3c}
\end{eqnarray}
where we again use strongly correlated noises $\xi_1=\xi_2=\xi$ with $D_{11}=
D_{22}=D_{12}\equiv D$, while we assume $\xi_3 = \eta$ is uncorrelated with
$\xi$, so $D_{13}=D_{23}=0$ and $D_{33} \equiv D_{3}$. The Fokker-Planck
equations corresponding to Eqs.\ (\ref{eq1c}) and (\ref{eq3c}) are then:
\begin{eqnarray}
{\partial_t p}
&=& -{\partial_x} \left[\gamma x (1-\epsilon x)p \right]
 + D \partial_x \Bigl[x(1-\epsilon x) \partial_x [x(1-\epsilon x) p]\Bigr]
+ D_3 \partial_{xx} p,
\label{eq201}\\
{\partial_t p}
&=& -{\partial_y} \left[\gamma y(1-\epsilon \ln{y})p \right]
 + D \partial_y \Bigl[y(1-\epsilon \ln{y}) \partial_y [y(1-\epsilon \ln{y}) p]\Bigr]
+ D_3 \partial_{yy} p.
\label{eq2010}
\end{eqnarray}
Because the second-derivative terms are now strictly positive, diffusion will
prevent infinitely sharp gradients from forming, and the solutions will instead
ultimately equilibrate to stationary distributions. One key question is then to
what extent the previous unimodal versus bimodal behaviour remains the same
once $D_3$ is added to the problem.

\subsection{Logistic stationary distribution}
For the logistic model it is possible to derive the following analytic
expression for the stationary solutions to Eq.\ (\ref{eq201}):
\begin{eqnarray}
p(x)&=& \frac{1}{\sqrt{D(x-\epsilon x^2)^2 + D_3 }}\,\exp{\left\{\frac{\gamma}{4 D}
\left[\frac{1}{c_1}\ln{\left|\frac{z+{c_1}/{\epsilon}}{z-{c_1}/{\epsilon}}\right| }
    + \frac{1}{c_2}\ln{\left|\frac{z+{c_2}/{\epsilon}}{z-{c_2}/{\epsilon}}\right|}
 \right]\right\}},
\nonumber\\
&=& \frac{1}{\sqrt{D(x-\epsilon x^2)^2 + D_3 }}\,
{\left|\frac{z+{c_1}/{\epsilon}}{z-{c_1}/{\epsilon}}\right|}^{{\gamma}/{4 c_1 D}}\,
{\left|\frac{z+{c_2}/{\epsilon}}{z-{c_2}/{\epsilon}}\right|}^{{\gamma}/{4 c_2 D}},
\label{eq202}
\end{eqnarray}
where 
\begin{equation}
z=x-\frac{1}{2 \epsilon},\,\, 
c_1 = \sqrt{ i \epsilon \alpha + 1/4},\,\, 
c_2 = \sqrt{- i \epsilon \alpha + 1/4},\,\, 
\alpha = \sqrt{D_3/D}.
\label{eq203}
\end{equation}
See Appendix B for the details of this derivation. Figure 6 shows examples of
these solutions, for $D=10$ and $D_3=10^{-2}$ to $10^{-5}$. It is gratifying
to note that the solutions are still bimodal, so this feature is preserved.
Also, as one might expect, smaller $D_3$ yields peaks, at both the stable point
$e$ and the unstable point 0, that are both narrower and higher. It is worth
explicitly noting though that the case without an additive noise ($D_3=0$)
cannot be obtained by naively taking the limit of $D_3 \to 0$ in
Eq.\ (\ref{eq202}), as $D_3 \to 0$ is a singular limit \cite{SCHENZLE79}.
Without additive noise stationary distributions simply do not exist, and the
time-dependent PDF is given by Eq.\ (\ref{eq6}).
This is similar to the impossibility of recovering the Euler equation for the
inviscid fluid from the Navier-Stokes equation for the viscous fluid by taking
the limit of zero viscosity.

\begin{figure}
\centering
\includegraphics[scale=0.9]{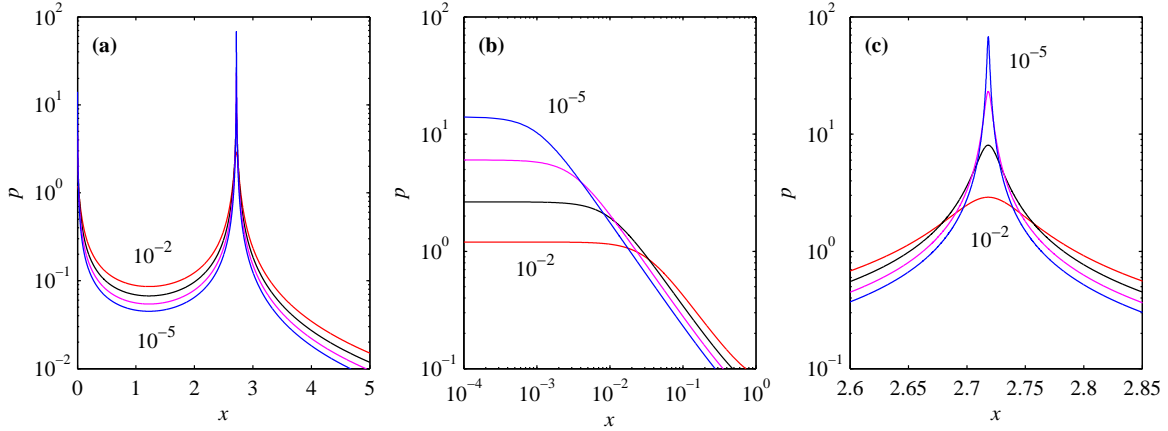}
\caption{All three panels show the same quantity, namely the logistic
stationary distribution given by Eq.\ (\ref{eq202}), with $D=10$, and $D_3=
10^{-2}$ to $10^{-5}$ as indicated. (a) shows the entire structure; note
incidentally how $p(x)$ now extends into the region $x>e$ as well. (b) shows
the details of the peak at 0, and (c) shows the peak at $e$.}
\vskip 1cm
\label{Fig6}
\end{figure}

\subsection{Numerical Solutions}
For the Gompertz model the integrals that would be required to obtain the
stationary solutions to Eq.\ (\ref{eq2010}) cannot be evaluated analytically.
Also, if full time-dependent solutions are desired, then neither
Eq.\ (\ref{eq201}) nor (\ref{eq2010}) have analytic solutions. Numerical
solvers to both the steady-state and time-dependent problems were developed,
based on standard finite-difference formulas. The results are second-order
accurate in space, and also in time for the time-stepping version. Most aspects
of these solvers are standard, so details are not presented. The only aspect
that requires discussion are the associated boundary conditions. We begin by
summarizing either of Eq.\ (\ref{eq201}) or (\ref{eq2010}) in the generic form
\begin{equation}
\partial_t p = -\partial_z[F(z)p] + D\partial_z[G(z)\partial_z(G(z)p)]
+D_3\partial_{zz}p,
\label{genericform}
\end{equation}
where $z$ represents either $x$ or $y$, and the functions $F$ and $G$ are as
appropriately defined for the two models. If we then integrate
Eq.\ (\ref{genericform}) between $z=0$ and some upper boundary $Z$, we obtain
\begin{equation}
\frac{d}{dt}\,\int_0^Z p\,dz
= (-Fp + D\,G\partial_z(Gp) + D_3\partial_z p)\Big|_0^Z.
\end{equation}
Now, $F(0)=G(0)=0$ for both logistic and Gompertz models, so this becomes
\begin{equation}
\frac{d}{dt}\,\int_0^Z p\,dz
= (-Fp + D\,G\partial_z(Gp) + D_3\partial_z p)(Z) - D_3\partial_z p(0).
\label{boundary}
\end{equation}
Certainly in the limit $Z\to\infty$ we require that the total probability
integral should remain constant, so the boundary condition at $z=0$ must be
that $\partial_z p=0$. Of course, $Z\to\infty$ cannot be achieved in any
numerical solver; some finite upper boundary must always be chosen. In
previous work on other Fokker-Planck equations \cite{KH17,HK17},
the resulting PDFs dropped off sufficiently rapidly for large $z$
(exponentially or even faster) that just imposing $p(Z)=0$ yielded excellent
results, and conserved the total probability extremely well. Here though this
approach was found not to work, and caused the integral $\int_0^Z p\,dz$ to
decrease in time, even if $Z$ as large as 100 was chosen. The reason
is that here $p$ decreases so slowly ($\sim1/z^2$) for large $z$ that
unacceptably large values of $Z$ would have to be chosen to make $p(Z)$
sufficiently small for $p(Z)=0$ to be a reasonable approximation. Fortunately,
Eq.\ (\ref{boundary}) already provides the remedy: if the outer boundary
condition is simply chosen to be
\begin{equation}
-Fp + D\,G\partial_z(Gp)+ D_3\partial_z p=0\qquad{\rm at}\quad z=Z,
\label{boundcond}
\end{equation}
then taking $Z$ as small as 10 works very well, with the probability integral
properly conserved. Spatial grids up to $10^7$ grid points were used, and
results were carefully checked to ensure they were independent of the grid
size, time step, and precise choice of outer boundary $Z$.

\subsection{Diagnostics of stationary distributions}

The qualitative features of the Gompertz stationary distributions are the same
as previously seen in Figure 6 for the logistic model. Figure 7 summarizes how
the peaks and widths of the peaks behave in the two models, for $D$ varying
from 0.1 to 100, and $D_3=10^{-2}$ to $10^{-5}$. We see that the amplitude of
the main peak at $e$ is very similar in both models, hardly varies with $D$,
and scales with $D_3$ as $D_3^{-1/2}$. The amplitude at 0 is very small
for $D<1$, but rises rapidly thereafter. For the Gompertz model $D=1$ is
already sufficient to have a local maximum at the origin; for the logistic
model $D\approx1-2$ is required (depending on $D_3$). For $D=100$ both models
have peaks at the origin that are almost as large as the peaks at $e$. Unlike
the previous Gompertz singularity at the origin though, there is now no case
where the $0$-peak exceeds the $e$-peak.

Turning next to the widths (which we define to be the width at half the peak
amplitude), the variation with $D_3$ is as one might expect, namely $\sim
D_3^{1/2}$, for all $D$. The variation with $D$ is less obvious, indeed
somewhat counter-intuitive. For $D\le O(1)$ the widths of the peaks at $e$
hardly vary with $D$, whereas for $D\ge O(1)$ they {\it decrease} as
$D^{-1/2}$. That is, even though it is larger $D$ which is causing the PDFs to
spread out from the stable equilibrium point, in the immediate vicinity
of $e$ a larger $D$ yields a narrower peak. The widths of the peaks at 0 show
a similar $D^{-1/2}$ scaling in the $D\ge O(1)$ regime where they are peaks at
all.

\begin{figure}
\centering
\includegraphics[scale=0.9]{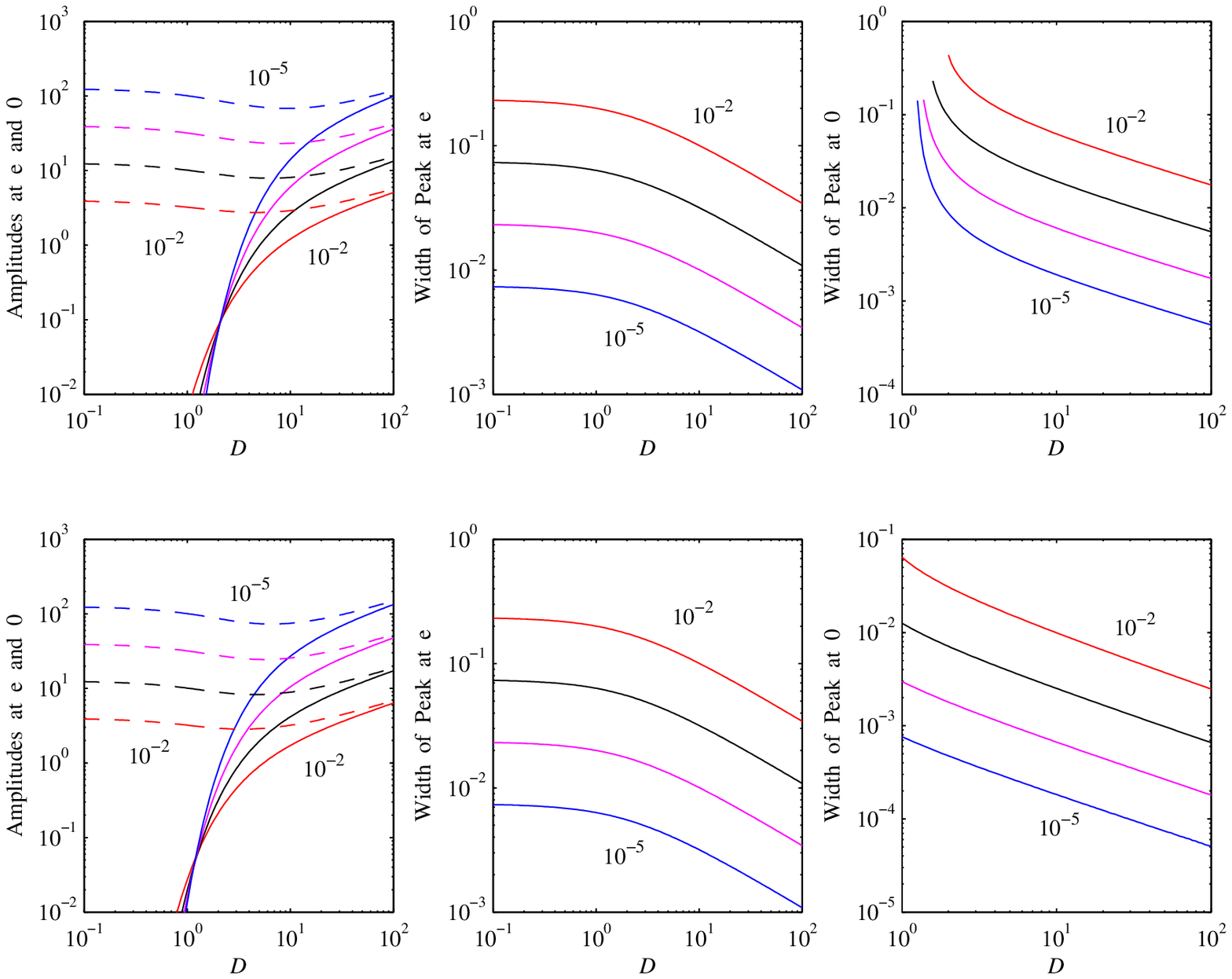}
\caption{The top row shows results for the logistic model, the bottom row
for the Gompertz model. The first panel in each row shows the amplitudes
$p(e)$ (dashed) and $p(0)$ (solid). $D_3=10^{-2}$ to $10^{-5}$ as indicated.
The second panel in each row shows the widths at half-peak of the peaks at $e$,
again for $D_3$ as indicated. The final panels show the widths at half-peak of
the peaks at 0. Note how $D$ here only covers a smaller range than in the
other panels; for smaller $D$ the origin is either not a peak at all, or not
yet sufficiently dominant to have a corresponding half-peak.}
\vskip 1cm
\label{Fig7}
\end{figure}

Figure 8 shows three further diagnostic quantities, all intended to measure
the extent to which a PDF is localised versus spread out. First we have the
familiar standard deviation $\sigma$, defined as
\begin{equation}
\sigma=\left(\int(z-\langle z\rangle)^2 p\,dz\right)^{1/2},\quad
\langle z\rangle=\int zp\,dz,
\end{equation}
where $z$ is either $x$ for logistic or $y$ for Gompertz. As long as $D$ is
sufficiently small that the PDFs are unimodal, $\sigma$ is a good measure of
localisation, measuring much the same (to within a multiplicative constant) as
the widths at half-peak. Once the PDFs become bimodal though, $\sigma$ is
largely useless, and is only measuring the distance between the two peaks
rather than any details associated with either peak. Next we have the
so-called differential entropy
\begin{equation}
S \propto -\int p\ln p\,dz,
\label{entropy}
\end{equation}
where the Boltzmann constant $K_B$ is not shown explicitly. Eq.\ (\ref{entropy})
is a measure of disorder and variability, and is thus expected to be small for
highly localised PDFs and large for spread out ones (e.g.\ 
\cite{fisher,KH17,HK17}). We see that entropy does a far better job than
$\sigma$ did of still distinguishing structures even in the bimodal regime;
note how $S$ continues to vary with both $D$ and $D_3$ even in the regime
where $\sigma$ has become useless. Finally, another useful measure of
information is the Fisher information
\begin{equation}
I=\int\frac{(\partial_z p)^2}{p}\,dz,
\end{equation}
which is expected to have the opposite behaviour as the entropy
(e.g.\ \cite{fisher,KH17,HK17}). We see that Fisher information again shows
variation with $D$ and $D_3$ even in the bimodal regime, where it also
distinguishes the most between the two models. We conclude therefore that the
most useful diagnostics of variability in bimodal structures are the Fisher
information and then entropy, while $\sigma$ is only capturing the distance
between peaks but nothing else about the PDFs.

\begin{figure}
\centering
\includegraphics[scale=0.9]{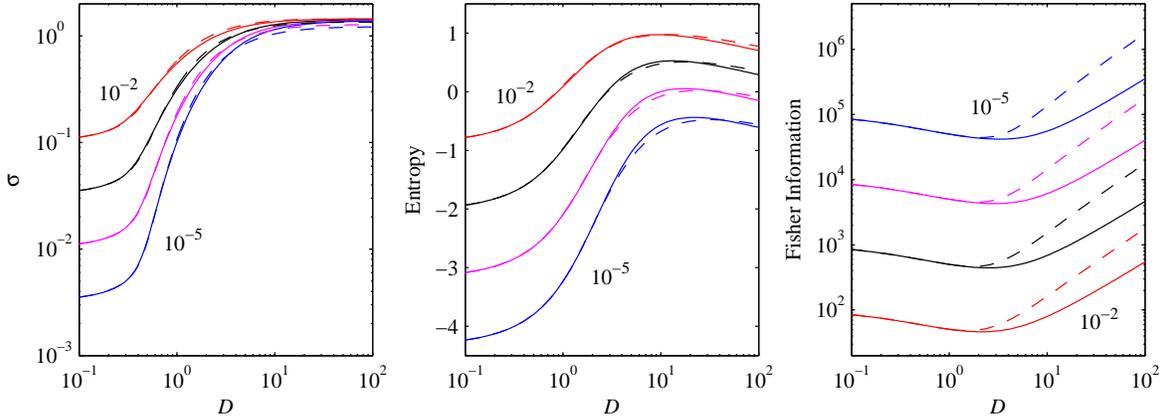}
\caption{From left to right plots of standard deviation $\sigma$, entropy
and Fisher information as functions of $D$, for $D_3=10^{-2}$ to $10^{-5}$
as indicated. Solid lines show results for the logistic stationary
distribution, dashed lines for Gompertz.}
\vskip 1cm
\label{Fig8}
\end{figure}

\subsection{Time-dependent solutions}

Finally, a substantial number of runs was done exploring how different
initial conditions evolve toward the stationary distributions considered
previously. It was found that logistic and Gompertz models behave similarly;
only logistic results will therefore be presented in detail here. We start
with the initial condition
\begin{equation}
p_0=\sqrt{\frac{500}{\pi}}\,exp\Bigl[-500(x-x_0)^2\Bigr],
\label{initcond}
\end{equation}
and varied $x_0$ in the range $(0.1,4)$. The factor 500 was chosen to make
the initial Gaussian peak slightly narrower but comparable to the expected
final distributions. Taking even narrower initial conditions simply
involves additional broadening, but otherwise qualitatively the same
behaviour. Similarly, $D_3=0.01$ is fixed here; other values were explored,
and yielded results generally similar, just with different widths as explored
before for the stationary distributions.\\

Figure 9 shows results for the total information length ${\cal L}_\infty$
that occurred when starting from this initial condition (\ref{initcond}) and
evolving the solution to the final stationary distribution. We see how
logistic and Gompertz models are indeed very similar. Both have
${\cal L}_\infty\approx O(10)$ for $D=0.1$, 1 and 10. The dip around
$x_0\approx 0$ reflects that the multiplicative noise endows the
certain degree of stability to the unstable equilibrium point (in the 
absence of the noise), making it more similar to the equilibrium
point $e$; if the initial peak is already
near its final position, then very little information change is needed to
reach the final position. Alternatively, this suggests that the multiplicative
noise induces fast switching between the stable and unstable equilibrium points.\\

\begin{figure}
\centering
\includegraphics[scale=0.9]{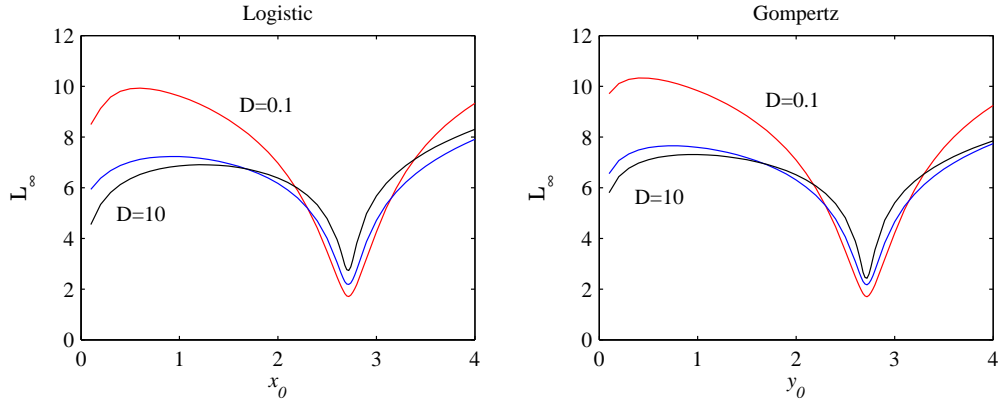}
\caption{${\cal L}_\infty$ as a function of the initial peak position, for
logistic on the left and Gompertz on the right. $D=0.1$, 1 and 10 as
indicated.}
\vskip 1cm
\label{Fig9}
\end{figure}

Figure 10 shows the detailed spatial structures throughout the evolution,
for the four representative cases $D=0.1$ and 10, and $x_0=0.5$ and 4.
Considering $D=0.1$ first, the solutions always remain relatively narrow,
as we might expect based on the previous results. The peaks move monotonically
from $x_0$ to $e$; that is, $x_0=0.5$ moves outward, and $x_0=4$ moves inward.
It is interesting to note though that in the intermediate stages there is
also a certain amount of diffusive spreading in the opposite direction. That
is, for $x_0=0.5$ the PDF in the region $x<0.5$ grows at least temporarily
(although never becoming the dominant peak), and similarly for $x_0=4$ the PDF
in the region $x>4$ grows temporarily.

For $D=10$ this movement away from the final position $e$ is even more
dramatic. For $x_0=0.5$ the peak itself moves toward 0, and it is only at
later times that a new peak at $e$ emerges and dominates. It was found that
all peaks with small $x_0$ initially move toward 0, whereas peaks with larger
$x_0$ immediately move toward $e$. The dividing line occurs near $x_0\approx
1$, where ${\cal L}_\infty$ in Figure 9 also has its local maximum. For
$x_0=4$ (and all $x_0>e$) the peak always moves toward $e$, but at least
temporarily there is also a very substantial contribution in the region $x>4$.
(These results were done with the computational outer boundary set to $Z=25$,
but thanks to the boundary condition (\ref{boundcond}), even $Z=10$ already
yields results that are essentially indistinguishable.)\\

\begin{figure}
\centering
\includegraphics[scale=0.9]{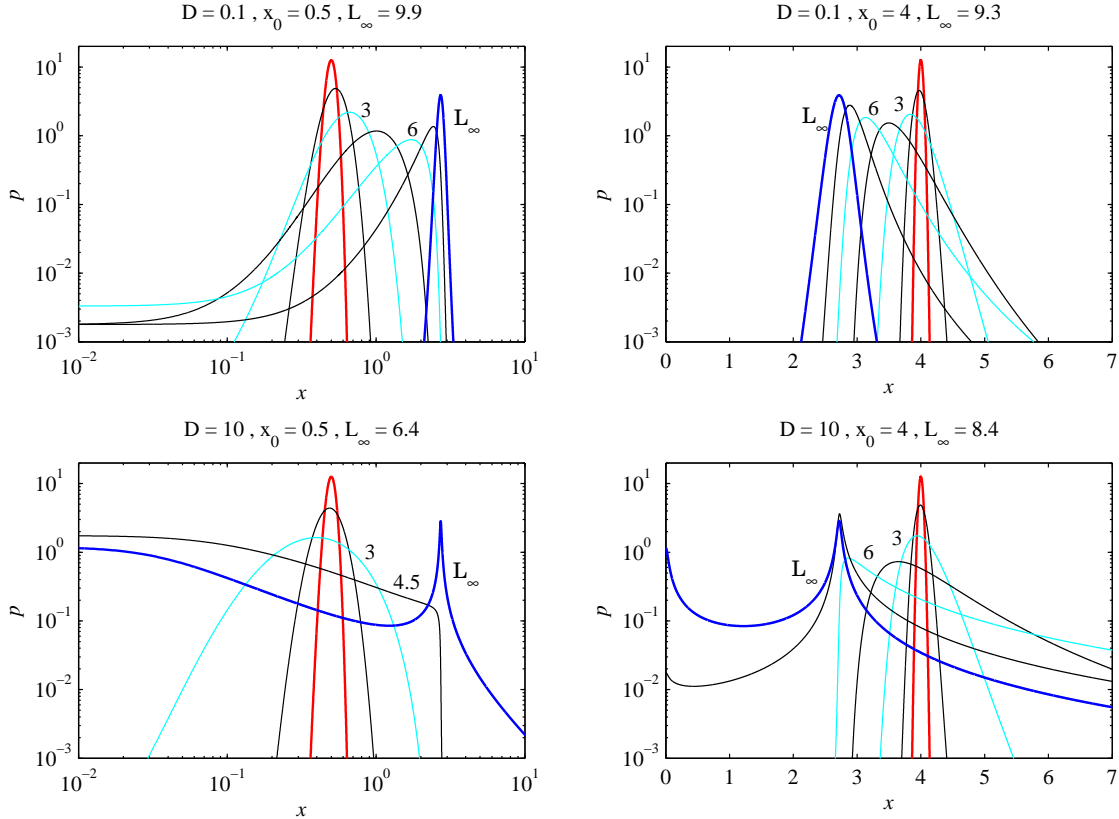}
\caption{Solutions of the logistic Fokker-Planck equation (\ref{eq201}),
with $D_3=0.01$ fixed, and $D=0.1$ and 10, and $x_0=0.5$ and 4 as indicated.
The initial condition (\ref{initcond}) is the thick line peaking at the given
$x_0$ value. The thick lines labelled ${\cal L}_\infty$ are the final
stationary distributions, the same as previously in Eq.\ (\ref{eq202}).
The thin lines intermediate between initial and final states were chosen to
have increments in ${\cal L}$ of 1.5, as indicated also by the numbers beside
some of them.}
\vskip 1cm
\label{Fig10}
\end{figure}

Figure 11 shows some diagnostic quantities that can be computed for
the time-dependent PDFs in Figure 10. In the top row we have two measures of
position, namely the position of the peak itself, and the average value
$\langle x\rangle=\int xp\,dx$. For $D=0.1$ both quantities behave much the
same, simply moving monotonically from $x_0$ to $e$. For $D=10$ they behave
quite differently. For $x_0=0.5$ the position of the peak is as we saw before
in Figure 10; that is, it moves toward 0, until eventually a new peak emerges
at $e$ and suddenly becomes the dominant peak. This abrupt transition
is related to the dip of ${\cal L}_\infty$ in Figure 9 for small $x_0$,
reflecting a sudden switching between 
the unstable and stable equilibrium points. In contrast, $\langle x\rangle$
still evolves monotonically toward $e$. For $x_0=4$ the position of the peak
moves monotonically toward $e$, again as seen in Figure 10. It is
$\langle x\rangle$ which now does something unexpected, namely initially
increase to values significantly greater than 4. The explanation of this is
the phenomenon we saw before in Figure 10, that in the intermediate stages
there is very significant diffusive spreading to the region $x>4$, and at
least initially the peak spreads far more toward $x>4$ than toward $x<4$.

Finally, the bottom row of Figure 11 shows the entropy (\ref{entropy}), which
is again seen to be a useful measure of how spread out the PDF is. In particular,
the initial conditions (\ref{initcond}) all start with $S\approx-2$, whereas
the final values (which are the same as the corresponding results in Figure 8),
are always greater, consistent with the fact that the final distributions are
indeed more spread out than the initial conditions. We see though that in
three of the four cases presented here, the entropy is not monotonic,
indicating that at intermediate stages of the evolution the PDFs are even
more spread out.

\begin{figure}
\centering
\includegraphics[scale=0.9]{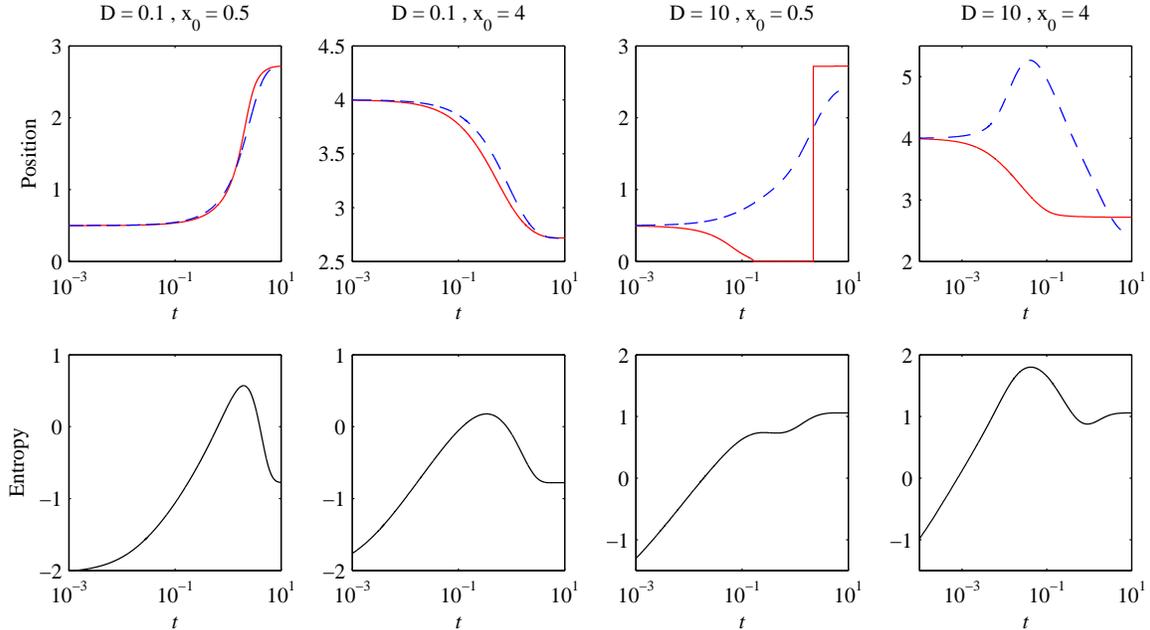}
\caption{The top row shows the position of the peak (solid) and the average
value $\langle x\rangle$ (dashed) as functions of time, for the four logistic
model solutions from Figure 10, as indicated. The bottom row shows
corresponding results for the entropy $S(t)$.}
\vskip 1cm
\label{Fig11}
\end{figure}

\section{Conclusions}
A growth model is widely used in understanding the dynamic evolution of
populations, and two of the most popular models are the logistic and Gompertz
models. The key difference between these two lies in different nonlinear
damping, which is weaker in the Gompertz model than a quadratic damping in
the logistic model. We examined consequences of this different nonlinear
damping in attractor structure and time-evolution of stochastic logistic and
Gompertz models by considering different types of stochastic noise. A
stochastic noise $D$ from a multiplicative (internal, e.g.\ epigenetic)
source was shown to induce an interesting transition from unimodal to bimodal
PDFs in both models as the attractor around the stable equilibrium point
becomes leaky. A much weaker damping in the Gompertz model led to a much more
significant (e.g.\ singular) growth of the population of a very small size
around the unstable equilibrium point. Time-dependent PDFs were shown to be
drastically different from stationary PDFs, while  the variance is not a
true representation of the variability of a bimodal PDF, highlighting the
importance of time-dependent PDFs. We also showed that the multiplicative
noise can induce an abrupt switching between the unstable and stable
equilibrium points.\\

These results can have an interesting
implication for understanding the role of variability in experimentally
observed bimodal distributions. For instance, comparing the distribution of
inflammatory genes (experimentally observed to be bimodal) with house-keeping
genes (experimentally observed to be unimodal) \cite{SHALEK14}, we can infer
that the variability in inflammatory genes is higher than that in
house-keeping genes. Furthermore, our results imply that the maintenance of
the population of small size is preferred as the stochastic component
(e.g.\ epigenetic noise) in the growth rate dominates over the constant growth
rate, as it happens in an environment very unfavourable for survival, for
instance due to antibiotics or drugs. Our stochastic Gompertz model may be
more relevant to the case of the extreme limit of a significant population of
small size in such a scenario.\\

We also presented the information geometry associated with a growth process
from the perspective of information change, and compare these two models in
terms of the information length which represents the total number of
statistically different states that the system undergoes in time. This
information length provides a useful system-independent method of analyzing
different stochastic processes to understand different experimental data. In
particular, we showed that even though the time-evolutions of the two models
are very different, they become more similar when measured in unit of the
information length. These results suggest an interesting utility of the
information length in unifying seemingly very different non-equilibrium
growth processes. 

\appendix
\section{Relation between ${\cal L}$ and relative entropy}
We first show the relation between $\tau(t)$ in Eq.~(\ref{eq201}) and the
second derivative of the relative entropy (or Kullback-Leibler divergence)
$D(p_{1},p_{2}) = \int dz \,p_{2} \ln{(p_{2}/p_{1})}$ where $p_{1}=p(z,t_{1})$
and $p_{2}=p(z,t_{2})$ as follows:
\begin{eqnarray}
\frac{\partial}{\partial t_{1}} D(p_{1},p_{2}) &=&
 - \int dz p_{2} \frac{ \partial_{t_{1}} p_{1}}{p_{1}},
\label{a1}\\
\frac{\partial^{2}}{\partial t_{1}^{2}}D(p_{1},p_{2}) &=&
 \int dz p_{2} \left[ \frac{ (\partial_{t_{1}} p_{1})^{2}}{p_{1}^{2} }
 - \frac{\partial_{t_{1}}^{2} p_{1}}{p_{1}} \right],
\label{a2}\\
\frac{\partial}{\partial t_{2}} D(p_{1},p_{2}) &=&
 \int dz \left[ \partial_{t_{2} }p_{2}
 + \partial_{t_{2}} p_{2} (\ln{p_{2}}-\ln{p_{1}}) \right ],
\label{a3}\\
\frac{\partial^{2}}{\partial t_{2}^{2}} D(p_{1},p_{2}) &=&
 \int dz \left[ \partial_{t_{2}}^{2} p_{2} 
 + \frac{(\partial_{t_{2}} p_{2})^{2}}{p_{2}}
 + \partial_{t_{2}}^{2} p_{2} (\ln{p_{2}}-\ln{p_{1}}) \right ].
\label{a4} 
\end{eqnarray}
By taking the limit where $t_{2} \to t_{1} = t$ ($p_{2} \to p_{1}=p$) and by
using the total probability conservation (e.g.\ $\partial_t\int dz p = 0$),
Eqs.~(\ref{a1}) and (\ref{a3}) above lead to
$$ \lim_{t_{2} \to t_{1}=t} \frac{\partial}{\partial t_{1}} D(p_{1},p_{2})
 =\lim_{t_{2} \to t_{1}=t} \frac{\partial}{\partial t_{2}} D(p_{1},p_{2})
 = \int dz \partial_{t } p=0,$$
while Eqs.~(\ref{a2}) and (\ref{a4}) give
$$ \lim_{t_{2} \to t_{1}=t} \frac{\partial^{2}}{\partial t_{1}^{2}}D(p_{1},p_{2})
= \lim_{t_{2} \to t_{1}=t} \frac{\partial^{2}}{\partial t_{2}^{2}} D(p_{1},p_{2})
= \int dz \frac{ (\partial_{t} p)^{2}}{p }.$$ \\

To link this to information length ${\cal L}$, we then express $D(p_{1},p_{2})$
for small $dt = t_2 - t_1$ as
\begin{equation}
D(p_{1},p_{2}) = \left[ \int dz
 \frac{ (\partial_{t_1} p(z,t_1))^{2}}{p } \right] (dt)^2 + O((dt)^3),
\label{a5}
\end{equation}
where $O((dt)^3)$ is higher order term in $dt$. We define the infinitesimal
distance (information length) $dl(t_1)$ between $t_1$ and $t_1 +dt$ by
\begin{equation}
dl(t_1) = \sqrt{ D(p_{1},p_{2})} = \sqrt{ \int dz
 \frac{ (\partial_{t} p)^{2}}{p }} dt + O((dt)^{3/2}).
\label{a6}
\end{equation}
The total change in information between time $0$ and $t$ is then obtained by
summing over $dt(t_1)$ and then taking the limit of $dt \to 0$ as
\begin{eqnarray}
{\cal L}(t) &= &\lim_{dt \to 0} \left [ dl(0) +dl(dt) +dl(2dt) + dl(3dt)
 + \cdot\cdot \cdot dl(t-dt) \right]
\nonumber\\
& = & 
\lim_{dt \to 0} \left [ \sqrt{ D(p(z,0),p(z,dt))} + \sqrt{ D(p(z,dt),p(z,2dt))}
 + \cdot\cdot\cdot \sqrt{ D(p(z,t-dt),p(z,t))} \right ]
\nonumber\\
& \propto& \int_0^t dt_1\, \sqrt{ \int dz \frac{ (\partial_{t_1} p)^{2}}{p }}.
\label{a7}
\end{eqnarray}

\section{Derivation of stationary solution (\ref{eq202})}
We look for the stationary solution of the Fokker-Planck
equation Eq.\ (\ref{eq201})
\begin{eqnarray}
0 &=& -\left[\gamma (x-\epsilon x^2)p \right]
 + D\Bigl[x(1-\epsilon x) \partial_x [x(1-\epsilon x) p]\Bigr]
+ D_3 \partial_x p.
\label{eqA1}
\end{eqnarray}
We define $G(x)=x(1-\epsilon x)$ and $F(x)=\gamma x (1-\epsilon x)$
and express Eq.\ (\ref{eqA1}) as
\begin{eqnarray}
0 &=& - Fp + DG\partial_x[Gp] + D_3 \partial_x p,\\
\partial_x p [D_3 + D G^2] &=& p [F-DG\partial_x G],\\
\partial_x p &=& p \frac{F}{D_3 + D_G^2}
 - \frac{1}{2}\frac{\partial_x [D_3 + DG^2]}{D_3 + DG^2}.
\end{eqnarray}
The integral over $x$ of the above equation gives us
\begin{equation}
p \propto \exp \left \{ 
\int_{x_0}^x \frac{\gamma}{D} \frac{x'(1-\epsilon x')}{\alpha ^2
 + x'^2(1-\epsilon x')^2} dx'
-\frac{1}{2} \ln \left[ D_3 + Dx^2(1-\epsilon x^2)\right]
\right \},
\label{eqA2}
\end{equation}
where $\alpha^2 = {D_3}/{D}$. In order to compute Eq.\ (\ref{eqA2}), we use
the partial fraction decomposition
\begin{equation}
\frac{\gamma}{D} \frac{x(1-\epsilon x)}{\alpha ^2 + x^2(1-\epsilon x)^2}
 = \sum_{i=1}^4 \frac{k_i}{x-\omega_i},
\label{eqA3}
\end{equation}
where $\omega_i$ are the complex solutions of
$\alpha ^2 + x^2(1-\epsilon x)^2$, which can be set as
\begin{equation}
\omega_i =\frac{1}{\epsilon} \left[ \frac{1}{2}
 \pm \sqrt{\frac{1}{4} \pm i \epsilon \alpha} \right ].
\end{equation}
In the following, we let
\begin{eqnarray}
c_1 = \sqrt{\frac{1}{4} + i \epsilon \alpha}, \qquad
c_2 = \sqrt{\frac{1}{4} - i \epsilon \alpha}, \qquad
y = x-\frac{1}{2 \epsilon}.
\end{eqnarray}
By using these notations, we compute the $k_i$ and Eq.\ (\ref{eqA3}) as
\begin{equation}
\frac{\gamma}{D} \frac{x(1-\epsilon x)}{\alpha ^2 + x^2(1-\epsilon x)^2} =
\frac{\gamma}{4 D c_1} \left[
\frac{1}{y + {c_1}/{\epsilon}}
-\frac{1}{y - {c_1}/{\epsilon}}
\right] + 
\frac{\gamma}{4 D c_2} \left[ 
\frac{1}{y + {c_2}/{\epsilon}}
-\frac{1}{y - {c_2}/{\epsilon}}
\right].
\label{eqA4}
\end{equation}
Thus, we obtain
\begin{eqnarray}
& &\int_{x_0}^x \frac{\gamma}{D}
 \frac{x'(1-\epsilon x')}{\alpha^2 + x'^2(1-\epsilon x')^2} dx'
\\
&=&\int_{x_0}^x 
\frac{\gamma}{4 D c_1} \left[
\frac{1}{y' + {c_1}/{\epsilon}}
-\frac{1}{y' - {c_1}/{\epsilon}}
\right] + 
\frac{\gamma}{4 D c_2} \left[ 
\frac{1}{y' + {c_2}/{\epsilon}}
-\frac{1}{y' - {c_2}/{\epsilon}}
\right] 
dx'\\
&=&\left [
\frac{\gamma}{4 D c_1} \ln 
\frac{y' + {c_1}/{\epsilon}}{y' - {c_1}/{\epsilon}} +
\frac{\gamma}{4 D c_2} \ln 
\frac{y' + {c_2}/{\epsilon}}{y' - {c_2}/{\epsilon}}
\right]_{x_0}^x,
\end{eqnarray}
which leads to Eq.\ (\ref{eq202}).

\end{document}